\documentclass[aps,prb,superscriptaddress,10pt]{revtex4-2}

\usepackage{amsmath} 
\usepackage{amssymb}
\usepackage{mathrsfs}
\usepackage{amsbsy}
\usepackage{bm}
\usepackage{graphicx}
\usepackage{times}
\usepackage{array}
\usepackage{color}
\usepackage{upgreek}
\usepackage{xcolor}
\usepackage{dsfont}
\usepackage{ulem}
\usepackage{textpos}
\usepackage{subfigure}

\newcommand{\eqr}[1]{Eq.~\eqref{#1}}
\newcommand{\hc}{^\dagger}

\graphicspath{{figures/}}

\begin{document}
\title{Two-photon absorption in semiconductors: a multi-band length-gauge analysis}

\author{W.-R.~Hannes}
\author{M.F.~Ciappina}
\affiliation{Physics Program, Guangdong Technion–Israel Institute of Technology, Shantou, Guangdong 515063, China}
\affiliation{Technion—Israel Institute of Technology, Haifa 32000, Israel}

\date{\today}

\begin{abstract}
The simplest approach to deal with light excitations in direct-gap semiconductors is to model them as a two-band system: one conduction and one valence band. 
For such models, particularly simple analytical expressions are known to exist for the optical response such as multi-photon absorption coefficients. 
Here we show that generic multi-band models do not require much more complicated expressions. 
Our length-gauge analysis is based on the semiconductors Bloch equations in the absence of all scattering processes.
In the evaluation we focus on two-photon excitation by a pump-probe scheme with possibly non-degenerate and arbitrarily polarized configurations.
The theory is validated by application to graphene and its bilayer, described by a tight-binding model, as well as bulk Zincblende semiconductors described by ${\bm {k\cdot}}{\bm p}$ theory.
\end{abstract}

\maketitle

\section{Introduction}

A broad variety of theoretical methods exists to investigate certain nonlinear susceptibilities of crystals. In the past, these methods have been often applied to simplified band-structure models, in many cases just two-band models \cite{Aversa1994,Hannes2019}. 
For certain materials like graphene \cite{Cheng2014} or hBN \cite{Attaccalite2018}, such models might capture the most relevant aspects of the optical response. However, for more complicated materials including Zincblende semiconductors, they are typically not sufficient for two reasons: (i) more than two states are optically coupled for the considered excitation frequency (e.g., due to valence band degeneracies) and (ii) the influence of further bands (like higher conduction bands) can be considerable, for instance by leading to anisotropies in the optical response~\cite{Hutchings1994,Dvorak1994}. 

The development of approaches to the nonlinear response of bulk semiconductors in more rigorous band-structure models has a long history, even though the respective results are not in agreement with each other in all details. 
For the case of Zincblende semiconductors, the Kane band-structure model has been used to evaluate two-photon absorption (2PA) coefficients based on the transition rate approach.~\cite{Bolger1993,Hutchings1992a,Dvorak1994} 
Murayami and Nakayama~\cite{Murayama1994,Murayama1995,Murayama1997} evaluated 2PA spectra using an \textit{ab initio} approach based on the microscopic response theory originally developed for atoms.~\cite{Boyd2008}
A more profound approach, which takes into account the subtleties of the dipole operator in an extended crystal, is the length gauge analysis by Aversa and Sipe.~\cite{Aversa1995}
This lead to general expressions for the non-linear susceptibility $\chi^{(3)}$-tensor in a multiband model. 
However, to our knowledge the expressions in Ref.~\cite{Aversa1995} have not yet been explicitly applied to any rigorous band-structure model. 
We can suspect two reasons: (i) due to the use of the length gauge they contain $\mathbf{k}$-derivatives of matrix elements, which are not directly available from numerical band-structure programs due to the random phase problem; (ii) The appearance of resonance denominators in these expressions renders a direct numerical evaluation of the sum over $\mathbf{k}$-values difficult, in particular in the limit of vanishing relaxation. 
Recently, Passos \textit{et al}~\cite{Passos2021} presented a similarly general analysis of the nonlinear optical conductivity in the independent particle approximation, which elucidated its structure as well as some general characteristics. However, the 2PA coefficients that could be derived from the presented (to be symmetrized) third-order conductivities still contain many terms, and the numerical evaluation was limited to rather simple models.

In this sense we can state that according to common belief, multi-photon absorption coefficients, like other nonlinear susceptibilities, are bulky expressions (after expanding commutators and performing differentiations), which, moreover, still need to be integrated over the relevant region of reciprocal space. 
Here we show that rather simple expressions are obtained with appropriate limitations. 
Our perturbative analysis is carried out in the length gauge as the previous works,~\cite{Aversa1995,Passos2021,Hannes2019} but is more specialized on multi (mostly two) -photon absorption coefficients in a pump-probe setting, where the probe delivers one photon and the pump the remaining photons to the absorption process.
Taking into account arbitrary (under the limitations) pump and probe frequencies and linear polarization directions (with respect to each other and the crystallographic axes), 
such experiments can reveal all relevant information about the nonlinear response (real part of optical conductivity) of crystals.
Since we consider the limit of absent scattering, the presented expressions for $\ell$PA coefficients are integrals over the  $\mathbf{k}$-points at which two states are optically coupled, i.e., where the corresponding interband transition is resonant with the total energy of $\ell$ absorbed photons. This means, that for a 3D crystal, only a 2D 
contour integral is to be evaluated over the resonance isosurfaces, which greatly reduces numerical cost. 
The integrand contains only three factors:
(i) the square modulus of a function ${\cal G}^{(n)}_{vc\mathbf{k}}$, which depends on
the excitation parameters, 
on transition energies, transition dipole moments, as well as their derivatives with respect to $\mathbf{k}$ (along the pulse directions);
(ii) $(f_{v\mathbf{k}}-f_{c\mathbf{k}})$, the difference in band populations, which thus do not enter in a more complicated way, e.g., via their $\mathbf{k}$-derivatives as one could expect~\cite{Passos2021};
(iii) a factor $\left\| \nabla_{\mathbf{k}} ({\omega}_{v\mathbf{k}}-{\omega}_{c\mathbf{k}})\right\|^{-1}$, the inverse norm of the transition group velocity.
In the linear order, ${\cal G}^{(1)}_{vc\mathbf{k}}$ is simply the transition dipole component along the probe polarization direction, while for $n>1$ ($n$ being odd), other transitions enter as well.

A different approach within the same length gauge framework is to model the response by dynamical simulations, that is by numerically solving the time evolution for finite-length pulses. While this has been carried out using a multiband parallel transport gauge in Ref.~\cite{KraussKodytek2021}, the computational effort is immense and increases quadratically with the number of modeled bands.
It also increases higher than linearly with the duration of the exciting pulses, since longer pulses require higher $\mathbf{k}$-resolution. For this reason, dynamical simulations are an alternative but not particularly convenient tool for the description of the nonlinear-optical response of crystals.
Since dynamical aspects do not play a crucial role in absorptive processes, 
a steady-state description is sufficient to even analyze the response to ultra-short pulses in the fs-range, as long as other processes (dephasing and relaxation) happen on longer time scales, which is typically the case in clean samples. 
For spectrally broad pulses, it might be necessary to take all frequency components of each pulse into account. In this work we include a comparison between the steady-state description, approximating each pulse by a single carrier frequency, with dynamical simulations, and find that these are in excellent agreement with each other.

Our article is organized as follows. In Sec.~\ref{sec:formalism} we derive the perturbative equations for the generic multiband model and relate the expectation value of the perturbative current with the optical conductivities / absorption coefficients. We also show the transformation of the $\mathbf{k}$-integral to the contour integral. Sec.~\ref{sec:evaluation} contains the explicit evaluation/simplificaiton of 1PA and 2PA coefficients in the general multiband model. The obtained results are applied in Sec.~\ref{sec:applications} to specific materials, namely mono- and bilayer graphene and Zincblende semiconductors. In the nearest-neighbour tight-binding model for monolayer graphene we obtain analytical results for both 1PA and 2PA. For GaAs and ZnSe, we study the  dependencies of 2PA on polarization and non-degeneracy parameter, and compare with dynamical simulations. We conclude in Sec.~\ref{sec:conclusions}.

\section{General formalism}
\label{sec:formalism}

\subsection{Perturbation theory} 
\label{ssec:Order-expansion}

For brevity, we derive the perturbative analysis directly from the semiconductor-Bloch equations (SBE) rather than the von Neumann equation. Scattering terms are neglected from the beginning.
Using the length gauge, the dynamical equation for the single-particle density matrix $\rho_{\mathbf{k}}$ is
\begin{align}\label{eq:SBE_length-gauge}
	\frac{\partial}{\partial t} \rho_{\mathbf{k}}
	&= 
	\frac{i}{\hbar} \big[ \epsilon_{\mathbf{k}} - e_0\bm{E}(t) \cdot 
	\bm \xi_{\mathbf{k}},\rho_{\mathbf{k}}\big]
	+
	\frac{1}{\hbar}  {e_0\bm{E}}(t) \cdot 
	{\nabla}_{\mathbf{k}} \rho_{\mathbf{k}}, 
\end{align}
where $\epsilon_{\mathbf{k}}$ is the diagonal matrix of band energies $\hbar\omega_{\lambda\mathbf{k}}$, and $\bm \xi_{\mathbf{k}}$ is the 
nonabelian Berry connection.
Since $\epsilon_{\mathbf{k}}$ is diagonal, we have restricted ourselves to proper gauges which is in contrast to the multiband gauge applied in Ref.~\cite{KraussKodytek2021}. 
The reason is that for non-diagonal  (gauge-transformed) $\epsilon_{\mathbf{k}}$ the matrix elements of $\rho_{\mathbf{k}}^{(n)}$ become coupled within each order, making an analytical solution rather involved.

The perturbative equations are obtained from the series $\rho_{\mathbf{k}}=\sum_{i=0}^\infty \rho_{\mathbf{k}}^{(i)}(t)$, where $\rho_{\mathbf{k}}^{(0)}=\rho_{\mathbf{k}}^{0}$ is the initial density matrix with $\rho_{\lambda\mu\mathbf{k}}^{0}=\delta_{\lambda\mu}f_{\lambda\mathbf{k}}$. For any $n>0$ we have
\begin{align}\label{eq:SBE_length-gauge_H}
	\frac{\partial}{\partial t} \rho_{\mathbf{k}}^{(n)}
	&= 
	i \omega_{\mathbf{k}} \circ \rho_{\mathbf{k}}^{(n)} - i \frac{e_0}{\hbar} {\bm{E}^{}}(t) \cdot 
	\left[\bm \xi_{\mathbf{k}},\rho_{\mathbf{k}}^{(n-1)}\right]
	+\frac{e_0}{\hbar}  {\bm{E}}(t) \cdot 
	{\nabla}_{\mathbf{k}} \rho_{\mathbf{k}}^{(n-1)},
\end{align}
where $\omega_{\lambda\mu\mathbf{k}} \equiv \omega_{\lambda\mathbf{k}}-\omega_{\mu\mathbf{k}}$, and $\circ$ is the elementwise product.
The electric field is Fourier transformed as
\begin{align}
	\label{eq:E-FT}
	\bm{E}(t)= \int \frac{\mathrm{d}\omega}{2\pi} { \bm{E}}_\omega e^{-i\omega t}
\end{align}
where the reality condition implies ${\bm{E}}_\omega={\bm{E}}_{-\omega}$.
The temporal dependence of the density matrix can also be written in the form of a Fourier transform
\begin{align}
	\label{eq:sol-form}
	\rho_{\mathbf{k}}^{(n)}(t)
	&= 
	\left(\frac{e_0}{2\pi\hbar} \right)^n
	\!\!
	\int\!\! \mathrm{d}\omega_1 \dots \mathrm{d}\omega_n
	e^{- i \omega_1 t} \dots e^{- i \omega_n t}  E_{\omega_1}^{\alpha_1} \dots E_{\omega_n}^{\alpha_n} {\cal P}_{\mathbf{k}}^{(n);\alpha_1\dots\alpha_n}
	,
\end{align}
with $\rho_{\mathbf{k}}^{(0)}={\cal P}_{\mathbf{k}}^{(0)}$. 
Here superscripts $\beta,\alpha_i$ indicate Cartesian components, which are to be summed over if repeated.
The tensor ${\cal P}_{\mathbf{k}}^{(n);\alpha_1\dots\alpha_n}$ depends on frequency arguments $(\omega_1,\dots,\omega_n)$, which have been dropped, since they can be inferred from the  Cartesian components.
Substitution into \eqr{eq:SBE_length-gauge_H} immediately gives the recursive solutions
\begin{align}
	\label{eq:sol-P}
	{\cal P}_{\mathbf{k}}^{(n);\alpha_1\dots\alpha_n} &= 
	\frac{1}{ \omega_1 + \dots + \omega_n +  \omega_{\mathbf{k}}} \circ
	\left( 
	i
	\partial^{\alpha_n} 
	{\cal P}_{\mathbf{k}}^{(n-1);\alpha_1\dots\alpha_{n-1}}  \right.
	\nonumber\\ &\qquad\qquad\qquad\qquad
	+
	\left.
	\left[ \xi_{\mathbf{k}}^{\alpha_n} , 
	{\cal P}_{\mathbf{k}}^{(n-1);\alpha_1\dots\alpha_{n-1}} \right]
	\right),
\end{align} 
where $\partial^{\alpha_i} \equiv {\partial}/{\partial k^{\alpha_i}} $

Instead of splitting the inhomogeneity in \eqr{eq:SBE_length-gauge} (and in the following analysis)
into the dipole matrix and the gradient terms, one could also divide it into an interband dipole matrix (with diagonal elements set to zero) and a generalized derivative (defined in \eqr{eq:gen_deriv}) term. 
This would correspond to a proper inter/intra-band term splitting; however, it does not seem to be beneficial for simplifying the analysis presented here.

\subsection{Macroscopic perturbative response}
\label{ssec:Current}

The macroscopic current density is given by
\begin{align}
	\bm{J} = -\frac{e_0}{L^3} \sum_{\mathbf{k}} \mathrm{Tr} ( \bm{v}_{\mathbf{k}} \rho_{\mathbf{k}} ),
\end{align}
where $L^3$ is the sample volume (including both confined and unconfined dimensions). 
The velocity matrix is
\begin{align}
	\label{eq:velocity}
	\hbar\bm{v}_{\mathbf{k}} &= {\nabla}_{\mathbf{k}} \epsilon_{\mathbf{k}} + i \big[\epsilon_{\mathbf{k}},\bm \xi_{\mathbf{k}}\big],
\end{align}
or
\begin{align}
	\label{eq:velocity_inter}
	\bm{v}_{\lambda\mu\mathbf{k}}^{} = 
	\left\{
	\begin{array}{ll}
		\nabla_{\mathbf{k}} \omega_{\lambda\mathbf{k}}^{}, &\lambda=\mu \\[8pt]
		i \omega_{\lambda\mu\mathbf{k}}^{} \bm{\xi}_{\lambda\mu\mathbf{k}}^{}, &\lambda\neq\mu \\
	\end{array}
	\right.
\end{align}
The perturbative response is obtained simply by substituting the perturbative density matrix. In particular,
\begin{align}
	\label{eq:Jn}
	J^{(n)\beta}(t) &= -\frac{e_0}{L^3} \sum_{\mathbf{k}} \mathrm{Tr} \left( v_{\mathbf{k}}^\beta \rho_{\mathbf{k}}^{(n)}(t) \right),
\end{align}
The relation between the applied fields and the current response is given by a conductivity tensor,
\begin{align}
	\label{eq:J_sigma_E}
	J^{(n)\beta}(t)
	&= 
	\int \frac{\mathrm{d}\omega_1 \dots \mathrm{d}\omega_n}{(2\pi)^n} 
	e^{- i \omega_t t} E_{\omega_1}^{\alpha_1} \dots E_{\omega_n}^{\alpha_n} 
	{\sigma}^{(n)}_{\beta\alpha_1\dots\alpha_n},
\end{align}
where $\omega_t=\omega_1 + \dots + \omega_n$.
Comparison with \eqref{eq:sol-form} and \eqref{eq:Jn} shows that the conductivity tensor is
\begin{equation}
	\label{eq:sigma-n}
	{\sigma}^{(n)}_{\beta\alpha_1\dots\alpha_n}
	=
	-\frac{e_0}{L^3} \left(\frac{e_0}{\hbar} \right)^n \sum_{\mathbf{k}} \mathrm{Tr} \left( v_{\mathbf{k}}^\beta {\cal P}_{\mathbf{k}}^{(n);\alpha_1\dots\alpha_n} \right).
\end{equation}
The arguments  $(\omega_t;\omega_1,\dots,\omega_n)$ have been again dropped for shortness. 
The symmetrization with respect to the $n$ components and corresponding frequency arguments is carried out specifically for the setting described in the next subseciton; a complete symmetrization at this point is not useful.

\subsection{Probe absorption induced by the pump}
\label{ssec:Absorption}

In this work we investigate multi-photon absorption in a pump-probe setting, where the pump and probe beam are assumed to be nearly co-propagating along the normal to the sample surface.
Following Ref.~\cite{Dvorak1994}, we denote the probe by 
${p}$ 
and the pump by ${e}$ (for `excitation').
The probe, for example, is characterized by intensity $I_{p}$, frequency $\omega_{p}$, polarization direction $\hat{\mathbf{p}}$, and propagation direction $\hat{\bm{k}}_{p}$.  
In this setting, there are typically multiple absorption processes scaling with different powers of $I_{p}$ and $I_{e}$; here, when analyzing $\ell$PA with some $\ell\ge 1$ we exclusively consider the process proportional to $I_{p} I_{e}^{\ell-1}$, i.e., where the probe delivers one and the pump the remaining photons of the absorption process. 
The total intensity change of $I_{p}$ over the propagation coordinate is made up of all processes; however, the coefficients for the other processes can be deduced from the one calculated here. For example, 
one-beam absorption can be easily deduced as a special case from the two-beam results presented here, as will become clear below. 
Two-beam experiments with linearly-polarized light are sufficient to completely investigate the third-order conductivity tensor.~\cite{Dvorak1994}

Even with the above restriction, there are different processes in the considered order ($\sim I_{p} I_{e}^{\ell-1}$).
In the language of the resonance-based analysis of Ref.~\cite{Passos2021}, we consider only $\ell=(n+1)/2$-photon resonant terms when analyzing the $n$-th order response. 
This means, only field products leading to a resonance denominator for $\omega_{p}+(\ell-1)\omega_{e}$ are considered. 
This is justified mainly by the fact that, typically, the lower-number photon absorption is off-resonant and thus vanishing in the assumed limit of absent scattering.
For gapless materials like graphene, the off-resonance condition of lower number photons cannot be met; however, the other response typically stems from a different set of $\mathbf{k}$-points and does not directly interact with $\ell$PA.
By this assumption we also exclude from our analysis other nonlinear-optical effects such as photovoltaic effects, DC field-induced second order response, and electro-optic effects.

The probe absorption in this $\ell$PA process is determined by the $n$-th order current response propagating in the probe direction, where $n-1$ out of $n$ field components belong to the pump. 
Here and in the following, we implicitly assume $\ell=(n+1)/2$.
For the `true' $\ell$PA as defined above, the positive frequencies come in the lowest orders ($i=1,\dots,\ell$), followed by the negative frequencies ($i=\ell+1,\dots ,2\ell-1$); the sum is $\omega_t=\omega_{p}$. 
For example, for 2PA ($n=3$), the combinations are ${\cal E}_p {\cal E}_e {\cal E}_e^*$ and ${\cal E}_e {\cal E}_p {\cal E}_e^*$,~\cite{Dvorak1994} where the field amplitudes ${\cal E}_p$ are defined as in~\ref{sec:absorption_pref}.  
For arbitrary $n$, the effective conductivity determining the absorption coefficient $\alpha^{(n)}_{p\{e\}}$ is thus
\begin{align}
	\label{eq:sigma_n}
	\sigma^{(n)}_\mathrm{eff} &= 
	\sum_{\cal P}
	{\sigma}^{(n)}_{p{\cal P}(p\{e\})\{e\}}(\omega_{p};{\cal P}(\omega_{p},\{\omega_{e}\}),\{-\omega_{e}\}) ,
\end{align} 
where ${\cal P}$ stands for permutations, and $\{\}$ for $(\ell-1)$-fold repetition of an element in a list.
We have also directly substituted the pump/probe indices in place of the cartesian indices; for instance,
${\sigma}^{(1)}_{p}\equiv {\sigma}^{(1)}_{\beta} p_\beta$.
This convention will be used throughout also for $\mathbf{k}$-dependent quantities such as
$\bm{v}_{\mathbf{k}}$, ${\cal P}_{\mathbf{k}}^{(n);\alpha_1\dots\alpha_n}$ etc.
It will be also convenient to introduce a symmetrized (over relevant terms) form of ${\cal P}_{\mathbf{k}}^{(n)}$ as
$\bar{\cal P}_{\mathbf{k}}^{(n)} \equiv \sum_{\cal P} {\cal P}_{\mathbf{k}}^{(n);p{\cal P}(p\{e\})\{e\}}$.
Using the slowly varying envelope approximation, in \ref{sec:absorption_pref}
it is shown that the relation between the conductivity tensor and the 
absorption coefficient is
\begin{equation}\label{eq:alpha_sigma_relation}
	\alpha^{(n)}_{p\{e\}} = K_{n} \mathrm{Re}\left[{\sigma}^{(n)}_\mathrm{eff}\right], \qquad
	K_{n} \equiv \frac{1}{n_{0{p}}n_{0{e}}^{(n-1)/2}}
	\frac{2}{(2\epsilon_0 c)^{(n+1)/2}},
\end{equation}
with background refractive indices, e.g.,  $n_{0{p}} \equiv n_0(\omega_{p})$. For degenerate and co-polarized pump and probe, the relation between the pump-probe and the single-beam $\ell$PA coefficient is
$\alpha^{(n)}_{p\{e\}} = \ell \alpha^{(n)}$. As mentioned above, the latter can thus be easily deduced from the more general pump-probe study. 
Substituting \eqr{eq:sigma-n}, we obtain
\begin{equation}
	\label{eq:alpha-ell}
	\alpha^{(n)}_{p\{e\}} =  
	- \tilde{K}_n
	\frac{1}{L^3}
	\sum_{\mathbf{k}} 
	\mathrm{Re}\left[
	\mathrm{Tr} \left( v_{\mathbf{k}}^{p} \bar{\cal P}_{\mathbf{k}}^{(n)} \right)
	\right],
\end{equation}
with
$\tilde{K}_n \equiv e_0 K_n \left(e_0/\hbar\right)^{n}$. 
Remember that $\alpha^{(n)}_{p\{e\}}$ depends on the frequencies and the polarization directions of the probe and pump beams.

As a final note, the absorption coefficient $\alpha^{(n)}_{p\{e\}}$ is related here with an effective conductivity tensor ${\sigma}^{(n)}_\mathrm{eff}$, which in turn can be expressed in terms of usual tensor components ${\sigma}^{(n)}_{\beta\alpha_1\dots\alpha_n}$. 
The common practice would be to first compute the set of relevant tensor components.
However, using $\bar{\cal P}_{\mathbf{k}}^{(n)}$ we have the option to compute the absorption individually for each polarization setting, for example, oblique polarization directions with respect to crystal axes.  
This strategy emerged mainly because the present study is a follow-up of a previous work based on dynamical simulations~\cite{KraussKodytek2021}. 
It also allows us to validate the polarization dependence with the one following from the tensor symmetry for the crystal class of the studied material.

\subsection{$\mathbf{k}$-integration}
\label{ssec:k-integration}

The $\mathbf{k}$-summation in Eq.~\eqref{eq:alpha-ell} is replaced by an integration (of same dimensionality ${D}$ as the sample),
$	L^{-3} \sum_{\mathbf{k}} = 
\kappa_{D} \int \mathrm{d}^{D}k$,
with $\kappa_{D} \equiv (2\pi)^{-{D}} (L_\mathrm{c}^{{D}-3})$.
The factor $L_\mathrm{c}^{3-{D}}$ contains the sample extensions in the confined directions. In this way we obtain a bulk  absorption coefficient, with units/dimensions independent of $D$. Without the factor $L_\mathrm{c}^{3-{D}}$, e.g. $\kappa_{D} \equiv (2\pi)^{-{2}}$ for quasi-2D materials, a reduced dimensionality absorption coefficient is obtained, denoted by $\alpha^{(n)}_{D\mathrm{D}}$. In a thin-film approximation, where the relative intensity change $\mathrm{d}I_{p}/I_{p}$ over the confined sample dimension, along which the probe propagates, is small,  $\alpha^{(n)}_{2\mathrm{D}}$ corresponds to the absorption integrated over $z$, which is just $\mathrm{d}I_{p}/I_{p}$.

Using $\mathrm{Re}(c)=\mathrm{Im}(ic)$ (with $c$ a complex number) and possibly using partial integration in order to take out the resonance denominator from $\mathbf{k}$-derivatives, we can write \eqr{eq:alpha-ell} in the form
\begin{align}
	\label{eq:alpha_form}
	\alpha^{(n)}_{p\{e\}} &= -\tilde{K}_n\ 
	\kappa_{D} 
	\sum_{\lambda\mu}
	\int \mathrm{d}^{D}\mathbf{k}\ 
	\mathrm{Im} 
	\frac{{\cal F}^{(n)}_{\lambda\mu\mathbf{k}}}{ \omega_\Sigma - {\omega}_{\mu\lambda\mathbf{k}} } ,
\end{align}
where $\omega_\Sigma = \omega_{p} + (\ell-1) \omega_{e}$ is the excitation frequency sum.
In this work, we reserve the band index $\lambda$ for valence bands (VBs) and the index $\mu$ for conduction bands (CBs).
At some places, the symbols $\lambda$ and $\mu$ are interchanged by symbols $v$ (VB) and $c$ (CB), respectively. 
When further (VB/CB) band indices are required, we use the symbols $\eta$ and $\eta'$.

The function ${\cal F}^{(n)}_{\lambda\mu\mathbf{k}}$ does not contain further $\omega_\Sigma$-resonance denominators. 
It does contain other resonance denominators in general, which may lead to divergences. 
In the models / parameters which we consider, this is typically not problematic.
For example, for gapped semiconductors, if both excitation frequencies are below gap, the other resonance denominators are for inter-valence or inter-conduction band transitions and never vanish at the same $\mathbf{k}$-point as the 2PA resonance denominator.

In the case of simple band-structure models, like those which are isotropic around the band center, it is obvious that the $\mathbf{k}$-integration can be carried out analytically. This has been demonstrated, e.g., for the low-energy description of graphene \cite{Cheng2014,Passos2021} or for a simplistic parabolic model of direct-gap semiconductors.~\cite{Passos2021}
However, it is in fact generally possible to transform the ${D}$-dimensional integral into (a sum of) resonance surface integrals (typically one for each resonant transition) of dimension ${D}-1$. This greatly reduces the numerical expenses and will be the procedure followed here.

We first choose a particular direction, denoted by unit vector $\hat{\mathbf{k}}_\parallel$ (not to be confused with a propagation unit vector), 
along which we carry out the integration by means of the Sokhotski–Plemelj theorem. 
We thus write
$\int \mathrm{d}^{D}\mathbf{k} = \int \mathrm{d}^{{D}-1}\mathbf{k}_\perp \int \mathrm{d}k_\parallel $ and
then we transform the $k_\parallel$-integral to frequency space.

To this end we split it at each extremum of ${\omega}_{\mu\lambda\mathbf{k}}$ along the integration line, so that in each segment ${\omega}_{\mu\lambda\mathbf{k}}$ is monotonous.
We label the extrema locations, including the start and end point of the $k_\parallel$-integral, by $\bar{k}_j$.
\begin{align}
	\label{eq:kint_split}
	\int \mathrm{d}k_\parallel 
	&=
	\sum_{j}
	\int_{\bar{k}_j}^{\bar{k}_{j+1}} \mathrm{d}k_\parallel 
	\\
	&= 
	\sum_{j}
	\int_{\bar{\omega}_{\bar{k}_j}}^{\bar{\omega}_{\bar{k}_{j+1}}} \mathrm{d} {\omega}_{\mu\lambda\mathbf{k}}
	\left( \frac{\partial {\omega}_{\mu\lambda\mathbf{k}}}{\partial k_\parallel}	\right)^{\!\!-1} 
	\\
	&= 
	\sum_{j}
	\int_{\bar{\omega}_{j,\mathrm{min}}}^{\bar{\omega}_{j,\mathrm{max}}} \mathrm{d} {\omega}_{\mu\lambda\mathbf{k}}
	\left| \frac{\partial {\omega}_{\mu\lambda\mathbf{k}}}{\partial k_\parallel}	\right|^{-1} .
\end{align}
This introduces singularities whenever $\partial {\omega}_{\mu\lambda\mathbf{k}}/\partial k_\parallel$ vanishes, but this problem is relieved below by another transformation.
In the last step we have just reversed the direction of integration in the segments with $\bar{\omega}_{\bar{k}_j}>\bar{\omega}_{\bar{k}_{j+1}}$.
Thus,
\begin{align}
	\alpha^{(n)}_{p\{e\}} &= -\tilde{K}_n\ 
	\kappa_{D} 
	\sum_{\lambda\mu}
	\int \mathrm{d}^{{D}-1}\mathbf{k}_\perp
	\sum_{j}
	\int_{\bar{\omega}_{j,\mathrm{min}}}^{\bar{\omega}_{j,\mathrm{max}}} 
	\mathrm{d} {\omega}_{\mu\lambda\mathbf{k}}
	\left| \frac{\partial {\omega}_{\mu\lambda\mathbf{k}}}{\partial k_\parallel}	\right|^{-1}
	\!\!\!
	\mathrm{Im}
	\frac{{\cal F}^{(n)}_{\lambda\mu\mathbf{k}}}{ \omega_\Sigma - {\omega}_{\mu\lambda\mathbf{k}} } 
	\label{eq:alpha_intdomega}
\end{align}
Now we carry out the integral 
by adding a damping term $i \gamma_{vc}$ to the resonance denominator and taking the limit $\gamma_{vc}\to 0^+$, in order to apply the Sokhotski–Plemelj theorem.
\begin{equation}\label{key}
	\lim_{\gamma_{vc}\to 0^+} \int d {\omega}_{\mathbf{k}} \frac{{ g }_{\lambda\mu\mathbf{k}}}{ \omega_\Sigma - {\omega}_{\mathbf{k}} + i \gamma_{vc} } 
	= 
	\left.
	-i\pi { g }_{\lambda\mu\mathbf{k}} \right|_{ \omega_\Sigma = {\omega}_{\mathbf{k}} }
	+
	{\cal P} \int d {\omega}_{\mathbf{k}} \frac{{ g }_{\lambda\mu\mathbf{k}}}{ \omega_\Sigma - {\omega}_{\mathbf{k}} },
\end{equation}
where ${\mathcal {P}}$ denotes the Cauchy principal value.
If ${\cal F}^{(n)}_{\lambda\mu\mathbf{k}} \in\mathds{R}$, then, after taking the imaginary part, the second term vanishes. 
We will see that for $n=1$ and $3$ this is indeed the case.
The final result becomes
\begin{align}
	\alpha^{(n)}_{p\{e\}} 
	&= \pi \tilde{K}_n\ 
	\kappa_{D} 
	\sum_{\lambda\mu}
	\int \mathrm{d}^{{D}-1}\mathbf{k}_\perp
	\sum_{k_{\parallel,\mathrm{res}}}
	\left| \frac{\partial {\omega}_{\mu\lambda\mathbf{k}}}{\partial k_\parallel}	\right|^{-1}
	{\cal F}^{(n)}_{\lambda\mu\mathbf{k}}.
	\label{eq:alpha_intdkperp}
\end{align}
Here $k_{\parallel,\mathrm{res}}$ are the discrete values of $k_\parallel$ (for given $\mathbf{k}_\perp$) at which ${\omega}_{\mu\lambda\mathbf{k}}=\omega_\Sigma$.

The integral (for ${D}>1$) over the perpendicular $\mathbf{k}$-components, $\int \mathrm{d}^{{D}-1}\mathbf{k}_\perp \sum_{k_{\parallel,\mathrm{res}}}$, is not in a suitable form for the numerical evaluation, because some of the lines $\mathbf{k}_\perp=const.$ necessarily touch the resonance isosurface of ${\omega}_{\mu\lambda\mathbf{k}}$ (nearly) tangentially. At these points the integrand has a singularity due to the factor $\left| \partial {\omega}_{\mu\lambda\mathbf{k}} / \partial k_\parallel \right|^{-1}$, unless it is compensated by a zero of  the function ${\cal F}^{(n)}_{\mathbf{k}}$, but this is not the case except for specific models. Even though the integral converges, it cannot be numerically evaluated easily. Fortunately this singularity is removed by transforming the $\mathbf{k}_\perp$-integral to a path/surface integral over the resonance isosurface. 
The proof is given in \ref{sec:transform-k-integral}. 
With this transformation, the absorption coefficient (\ref{eq:alpha_intdkperp}) becomes
\begin{align}
	\alpha^{(n)}_{p\{e\}} 
	&= \pi \tilde{K}_n \kappa_{D} \ 
	\sum_{\lambda\mu} 
	\oint \mathrm{d} \mathbf{k}_{\mathrm{res}}
	\frac{{\cal F}^{(n)}_{\lambda\mu\mathbf{k}} }{
		\left\| \nabla_{\mathbf{k}} {\omega}_{\mu\lambda\mathbf{k}}\right\|}.
	\label{eq:alpha_isosurface}
\end{align}
The denominator does not vanish since we assume to have a well-defined (i.e., sharp) resonance contour. The equivalence between Eqs.~\eqref{eq:alpha_intdkperp} and \eqref{eq:alpha_isosurface} is trivial for ${D}=1$, where $\oint \mathrm{d} \mathbf{k}_{\mathrm{res}}$ becomes $\sum_{k_{\parallel,\mathrm{res}}}$.

\section{General evaluation}
\label{sec:evaluation}

\subsection{Linear absorption coefficient}

To obtain the linear absorption coefficient we need to substitute \eqr{eq:sol-P}  for $n=1$,
\begin{align}
	{\cal P}_{\mathbf{k}}^{(1);{p}} &= 
	\frac{1}{ \omega_{p} +  \omega_{\mathbf{k}}} \circ
	\left( 
	i
	\partial^{{p}} 
	{\cal P}_{\mathbf{k}}^{(0)}  +
	\left[ \xi_{\mathbf{k}}^{{p}} , 
	{\cal P}_{\mathbf{k}}^{(0)} \right]
	\right),
\end{align} 
into \eqr{eq:alpha-ell}, also for $n=1$:
\begin{align}\label{key}
	\alpha^{(1)} \equiv \alpha^{(1)}_{p}
	&=  
	- \tilde{K}_{1}\kappa_{D}
	\int \mathrm{d}^{D}\mathbf{k}\  
	\mathrm{Im}\left[
	\mathrm{Tr} \left( i v_{\mathbf{k}}^{p} \bar{\cal P}_{\mathbf{k}}^{(1)} \right)
	\right],
\end{align}
where $\bar{\cal P}_{\mathbf{k}}^{(1)}={\cal P}_{\mathbf{k}}^{(1);{p}}$.
If $\omega_{p}$ is larger than any inter-valence or inter-conduction band transitions, which we assume here, then
\begin{align}
	\alpha^{(1)} &= - \tilde{K}_1\ 
	\kappa_{D} 
	\sum_{\lambda\mu}
	\int \mathrm{d}^{D}\mathbf{k}\ 
	\mathrm{Im} 
	\frac{{\cal F}^{(1)}_{\lambda\mu\mathbf{k}}}{ {\omega}_{\lambda\mu\mathbf{k}} + \omega_{p} },
	\\
	\label{eq:F1}
	{\cal F}^{(1)}_{\lambda\mu\mathbf{k}} &=
	{\omega}_{\mu\lambda\mathbf{k}} \left| \xi_{\lambda\mu\mathbf{k}}^{{p}} \right|^2 \left( \rho_{\lambda\lambda\mathbf{k}}^0 - \rho_{\mu\mu\mathbf{k}}^0 \right)
	=
	{\omega}_{\mu\lambda\mathbf{k}} \left| \xi_{\lambda\mu\mathbf{k}}^{{p}} \right|^2
	f_{\lambda\mu\mathbf{k}},
\end{align}
with $\lambda\in v, \mu\in c$. 
Since ${\cal F}^{(1)}_{\mathbf{k}}$ is real, we may use the integration method from Sec.~\ref{ssec:k-integration}, by simply substituting into \eqr{eq:alpha_isosurface} to obtain
\begin{align}
	\alpha^{(1)} 
	&= \pi \tilde{K}_1 \kappa_{D} \ 
	\sum_{\lambda\mu} 
	\oint \mathrm{d} \mathbf{k}_{\mathrm{res}}
	\frac{{\cal F}^{(1)}_{\mathbf{k}} }
	{\left\| \nabla_{\mathbf{k}} {\omega}_{\mu\lambda\mathbf{k}}\right\|}
	\\
	&= \pi \omega_{p}  \tilde{K}_1 \kappa_{D} \ 
	\sum_{\lambda\mu} 
	\oint \mathrm{d} \mathbf{k}_{\mathrm{res}}
	\frac{ \left| \xi_{\lambda\mu\mathbf{k}}^{{p}} \right|^2 f_{\lambda\mu\mathbf{k}} }
	{\left\| \nabla_{\mathbf{k}} {\omega}_{\mu\lambda\mathbf{k}}\right\|},
\end{align}
which agrees with the Fermi's golden rule.

\subsection{2PA coefficient}

We now turn to 2PA in the order $n=3$. Due to the considered pump-probe setting we can study, for instance, both degenerate and non-degenerate 2PA, as well as co- or cross-polarized excitation. 
Equation \eqref{eq:alpha-ell} for $n=3$ reads
\begin{align}\label{eq:alpha2}
	\beta_{pe} \equiv
	\alpha^{(3)}_{pe}
	&=  
	- \tilde{K}_{2}\kappa_{D}
	\int \mathrm{d}^{D}\mathbf{k}\  
	\mathrm{Im}\left[
	\mathrm{Tr} \left( i v_{\mathbf{k}}^{p} \bar{\cal P}_{\mathbf{k}}^{(3)} \right)
	\right].
\end{align}
The symmetrized ${\cal P}$ to be substituted follows from \eqr{eq:sol-P} for $n=3$,
\begin{align}
	\label{eq:P3}
	\bar{\cal P}_{\mathbf{k}}^{(3)} &= 
	\frac{1}{ \omega_{p} +  \omega_{\mathbf{k}}} \circ
	\left( 
	i
	\partial^{{e}} 
	\bar{\cal P}_{\mathbf{k}}^{(2)}  +
	\left[ \xi_{\mathbf{k}}^{{e}} , 
	\bar{\cal P}_{\mathbf{k}}^{(2)} \right]
	\right),
\end{align} 
with $\bar{\cal P}_{\mathbf{k}}^{(2)} = {\cal P}_{\mathbf{k}}^{(2);{pe}} + {\cal P}_{\mathbf{k}}^{(2);{ep}}$. 
Substituting \eqr{eq:P3} into \eqref{eq:alpha2} and carrying out a partial integration (to take out the 2PA resonance denominator from the $\mathbf{k}$-derivative) in the first term gives
\begin{align}
	\label{eq:alpha3_P2}
	\alpha^{(3)}_{pe}
	&=  
	- \tilde{K}_{2}\kappa_{D}
	\int \mathrm{d}^{D}\mathbf{k}\  
	\mathrm{Im}
	\sum_{\lambda\mu}
	\left[
	\bar{\cal P}_{\lambda\mu\mathbf{k}}^{(2)}
	\partial^{{e}}
	{\cal V}_{\lambda\mu\mathbf{k}}^{p}
	+ i
	\left[ \xi_{\mathbf{k}}^{{e}} , 
	\bar{\cal P}_{\mathbf{k}}^{(2)} \right]_{\lambda\mu}
	{\cal V}_{\lambda\mu\mathbf{k}}^{p}
	\right],
	\\
	{\cal V}_{\lambda\mu\mathbf{k}}^{p} &= \frac{v_{\mu\lambda\mathbf{k}}^{p}}{ \omega_{p} +  \omega_{\lambda\mu\mathbf{k}}} .
\end{align}
In the first order we have
\begin{align}
	{\cal P}_{\eta\eta'\mathbf{k}}^{(1);{p}}
	&= \delta_{\eta\eta'}	\frac{i \partial^{{p}} f_{\eta\mathbf{k}}}{\omega_{p}}
	-
	\frac{ \xi_{\eta\eta'\mathbf{k}}^{p} f_{\eta\eta'\mathbf{k}} }{ \omega_{p}^{} + \omega_{\eta\eta'\mathbf{k}}}
	\equiv
	{\cal P}_{\eta\eta'\mathbf{k}}^{(1);{p}[tra]} 
	+
	{\cal P}_{\eta\eta'\mathbf{k}}^{(1);{p}[ter]} .
\end{align}
In the second order, only the $vc$-elements are relevant,
\begin{align}
	\label{eq:sol-pe}
	{\cal P}_{\lambda\mu\mathbf{k}}^{(2);{pe}} &=
	\left[
	i \partial^{{e}} 
	{\cal P}_{\lambda\mu\mathbf{k}}^{(1);{p}[ter]} 
	+
	\sum_{\eta}
	\left(
	\xi_{\lambda\eta\mathbf{k}}^{{e}}
	{\cal P}_{\eta\mu\mathbf{k}}^{(1);{p}[tra]}
	-
	{\cal P}_{\lambda\eta\mathbf{k}}^{(1);{p}[tra]}
	\xi_{\eta\mu\mathbf{k}}^{{e}}
	\right)
	+
	\right.\nonumber\\
	&\quad
	\left.
	\sum_{\eta\in v}
	\xi_{\lambda\eta\mathbf{k}}^{{e}}
	{\cal P}_{\eta\mu\mathbf{k}}^{(1);{p}[ter]}
	-
	\sum_{\eta\in c}
	\xi_{\eta\mu\mathbf{k}}^{{e}}
	{\cal P}_{\lambda\eta\mathbf{k}}^{(1);{p}[ter]}
	\right]
	\frac{1}{\omega_{p} + \omega_{e} +  \omega_{\lambda\mu\mathbf{k}}}
	.
\end{align}
The other term ${\cal P}_{\mathbf{k}}^{(2);{ep}}$ in the symmetric $\bar{\cal P}_{\mathbf{k}}^{(2)}$ is obtained by interchanging ${p}$ with ${e}$.
By substitution into \eqr{eq:alpha3_P2} and comparison with \eqr{eq:alpha_form} we can determine ${\cal F}^{(3)}_{\lambda\mu\mathbf{k}}$ as
\begin{align}
	{\cal F}^{(3)}_{\lambda\mu\mathbf{k}} &=
	\left(
	i\partial^{{e}} 
	{\cal P}_{\lambda\mu\mathbf{k}}^{(1);{p}[ter]} 
	+
	\sum_{\eta}
	\left(
	\xi_{\lambda\eta\mathbf{k}}^{{e}}
	{\cal P}_{\eta\mu\mathbf{k}}^{(1);{p}[tra]}
	-
	{\cal P}_{\lambda\eta\mathbf{k}}^{(1);{p}[tra]}
	\xi_{\eta\mu\mathbf{k}}^{{e}}
	\right)
	+
	\right.
	\nonumber\\
	&\qquad
	\left.
	\sum_{\eta\in v}
	\xi_{\lambda\eta\mathbf{k}}^{{e}}
	{\cal P}_{\eta\mu\mathbf{k}}^{(1);{p}[ter]}
	-
	\sum_{\eta\in c}
	\xi_{\eta\mu\mathbf{k}}^{{e}}
	{\cal P}_{\lambda\eta\mathbf{k}}^{(1);{p}[ter]} 
	+ \{ {p}\leftrightarrow{e}\}
	\right)
	\nonumber\\
	&\quad
	\left(
	\partial^{{e}}
	{\cal V}_{\lambda\mu\mathbf{k}}^{p}
	+
	i
	\sum_\eta
	\left(
	\xi_{\eta\lambda\mathbf{k}}^{{e}}
	{\cal V}_{\eta\mu\mathbf{k}}^{p}
	-\xi_{\mu\eta\mathbf{k}}^{{e}}
	{\cal V}_{\lambda\eta\mathbf{k}}^{p}
	\right)
	\right).
\end{align}
Note that there are poles in the elements of the matrices ${\cal V}_{\mathbf{k}}^{p}$ and ${\cal P}_{\mathbf{k}}^\mathrm{(1)}$.
For the terms into which their $vc$ elements enter they result in singularities when one excitation frequency approaches zero. This is to be expected because all electron scattering processes are neglected.
On the other hand, when $vv$ or $cc$ elements enter, there are singularities when one excitation frequency matches an inter-valence or inter-conduction band transition. 
Since ${\cal F}^{(3)}_{\lambda\mu\mathbf{k}}$ is to be evaluated at 2PA resonant $\mathbf{k}$-points only, these poles are only problematic if $\omega_{e}$ is resonant with a $vc$ transition at the same point. 
In this case the analysis in Sec.~\ref{ssec:k-integration} would not hold due to the presence of multiple poles. However, for the models studied here, this is not problematic. For the gapped semiconductors we will consider each frequency to be smaller than the band gap. For bilayer graphene (Sec.~\ref{sssec:Bilayer}), a gapless multiband model, it can be problematic, but only at discrete values of excitation frequencies, where the singularities occur, so it is still safe to use this approach for our purposes.

We now proceed with the algebraic simplification of ${\cal F}^{(3)}_{\lambda\mu\mathbf{k}}$.
As long as $\omega_{\mu\lambda\mathbf{k}}$ occurs outside of any $\mathbf{k}$-derivative, it may be replaced by $\omega_\Sigma$.
The terms in the first factor of ${\cal F}^{(3)}_{\lambda\mu\mathbf{k}}$ can be evaluated as follows:
\begin{align}\label{key}
	i\partial^{{e}} 
	{\cal P}_{\lambda\mu\mathbf{k}}^{(1);{p}[ter]}  = 
	i \frac{ \xi_{\lambda\mu\mathbf{k}}^{{p}} \partial^{{e}} f_{\lambda\mu\mathbf{k}}
		}{\omega_{e}}
	+
	i \frac{ f_{\lambda\mu\mathbf{k}} \partial^{{e}} \xi_{\lambda\mu\mathbf{k}}^{{p}}
	}{\omega_{e}}
	+
	i\frac{\xi_{\lambda\mu\mathbf{k}}^{{p}}f_{\lambda\mu\mathbf{k}}\partial^{{e}}\omega_{\lambda\mu\mathbf{k}} }{\omega_{e}^2}  
	\\
	\sum_{\eta}
	\left(
	\xi_{\lambda\eta\mathbf{k}}^{{e}}
	{\cal P}_{\eta\mu\mathbf{k}}^{(1);p[tra]}
	-
	{\cal P}_{\lambda\eta\mathbf{k}}^{(1);p[tra]}
	\xi_{\eta\mu\mathbf{k}}^{{e}}
	\right)
	= -i \xi_{\lambda\mu\mathbf{k}}^{{e}}
	\frac{ \partial^{{p}} f_{\lambda\mu\mathbf{k}}}{\omega_{p}}
	\\	
	\sum_{\eta\in v}
	\xi_{\lambda\eta\mathbf{k}}^{{e}}
	{\cal P}_{\eta\mu\mathbf{k}}^{(1);p} =
	-\sum_{\eta\in v}
	\frac{\xi_{\lambda\eta\mathbf{k}}^{{e}}\xi_{\eta\mu\mathbf{k}}^{{p}} f_{\eta\mu\mathbf{k}} }{ \omega_{p} + \omega_{\eta\mu\mathbf{k}} }
	\\
	-\sum_{\eta\in c}
	\xi_{\eta\mu\mathbf{k}}^{{e}}
	{\cal P}_{\lambda\eta\mathbf{k}}^{(1);p} =
	\sum_{\eta\in c}
	\frac{f_{\lambda\eta\mathbf{k}}\xi_{\lambda\eta\mathbf{k}}^{{p}}\xi_{\eta\mu\mathbf{k}}^{{e}}}{ \omega_{p} + \omega_{\lambda\eta\mathbf{k}} }
	=
	-\sum_{\eta\in c}
	\frac{f_{\lambda\eta\mathbf{k}}\xi_{\lambda\eta\mathbf{k}}^{{p}}\xi_{\eta\mu\mathbf{k}}^{{e}}}{ \omega_{e} + \omega_{\eta\mu\mathbf{k}} }.
\end{align}
The terms in the second factor of ${\cal F}^{(3)}_{\lambda\mu\mathbf{k}}$ can be evaluated as follows:
\begin{align}\label{key}
	-i 
	\partial^{{e}}{\cal V}_{\lambda\mu\mathbf{k}}^{p} &= \omega_{p} 
	\left[
	\frac{1}{\omega_{e}^2}
	\xi_{\mu\lambda\mathbf{k}}^{{p}}
	\partial^{{e}} \omega_{\mu\lambda\mathbf{k}} 
	- \left(\frac{1}{\omega_{p}}+\frac{1}{\omega_{e}}\right) \partial^{{e}} \xi_{\mu\lambda\mathbf{k}}^{{p}}
	\right]
	\\
	\sum_{\eta=\lambda,\mu} \Big(\dots\Big) &=
	\omega_{p} 
	\left[
	\frac{1}{\omega_{p}^2}
	\xi_{\mu\lambda\mathbf{k}}^{{e}}
	\partial^{{p}} \omega_{\mu\lambda\mathbf{k}} 
	- i \left( \xi_{\lambda\lambda\mathbf{k}}^{{e}} - \xi_{\mu\mu\mathbf{k}}^{{e}} \right) \left(\frac{1}{\omega_{p}}+\frac{1}{\omega_{e}}\right) \xi_{\mu\lambda\mathbf{k}}^{{p}}
	\right]
	\\
	\sum_{\eta\notin\{\lambda,\mu\}} \Big(\dots\Big) &=
	{\xi}_{\mu\lambda\mathbf{k};{e}}^{p}
	-
	{\xi}_{\mu\lambda\mathbf{k};{p}}^{e}
	+ i \omega_{p} \sum_{\eta\notin\{\lambda,\mu\}}
	\left(
	\frac{\xi_{\mu\eta\mathbf{k}}^{{p}}\xi_{\eta\lambda\mathbf{k}}^{{e}}}{ \omega_{p} + \omega_{\eta\mu\mathbf{k}} }
	+ \{ {p}\leftrightarrow{e}\}
	\right),
\end{align}
where we made use of the following sum rule derived by Aversa and Sipe~\cite{Aversa1995}:
\begin{equation}\label{eq:sum-rule}
	{\xi}_{\mu\lambda\mathbf{k};{p}}^{e}
	-
	{\xi}_{\mu\lambda\mathbf{k};{e}}^{p}
	=
	i \sum_{\eta\notin\{\lambda,\mu\}}
	\left(
	\xi_{\mu\eta\mathbf{k}}^{{p}}
	\xi_{\eta\lambda\mathbf{k}}^{e}
	-\xi_{\mu\eta\mathbf{k}}^{{e}}
	\xi_{\eta\lambda\mathbf{k}}^{p}
	\right),
\end{equation}
along with the generalized derivative notation~\cite{Aversa1995}
\begin{align}
	\label{eq:gen_deriv}
	{\xi}_{\mu\lambda\mathbf{k};{\alpha_1}}^{\alpha_2} 
	&=
	\partial^{\alpha_1} {\xi}_{\mu\lambda\mathbf{k}}^{\alpha_2} - i \left( \xi_{\mu\mu\mathbf{k}}^{\alpha_1} - \xi_{\lambda\lambda\mathbf{k}}^{\alpha_1} \right) {\xi}_{\mu\lambda\mathbf{k}}^{\alpha_2}.
\end{align}
Applying these steps we obtain
\begin{align}
	{\cal F}^{(3)}_{\lambda\mu\mathbf{k}} &=
	\omega_{p} 
	f_{\lambda\mu\mathbf{k}} 
	\left[
	\left(
	\frac{\partial^{{e}}\xi_{\lambda\mu\mathbf{k}}^{{p}}
	}{\omega_{e}}
	+\frac{\xi_{\lambda\mu\mathbf{k}}^{{p}}\partial^{{e}}\omega_{\lambda\mu\mathbf{k}} }{\omega_{e}^2}  
	+i
	\sum_{\eta}
	\frac{\xi_{\lambda\eta\mathbf{k}}^{{e}}\xi_{\eta\mu\mathbf{k}}^{{p}}}{ \omega_{p} + \omega_{\eta\mu\mathbf{k}} }
	\right)
	+ \{ {p}\leftrightarrow{e}\}
	\right]
	 \nonumber\\
	&\quad\times
	\left[
	\left(
	\frac{\xi_{\mu\lambda\mathbf{k};{e}}^{{p}}}{\omega_{e}}
	-
	\frac{\xi_{\mu\lambda\mathbf{k}}^{{p}}
	\partial^{{e}} \omega_{\mu\lambda\mathbf{k}} }{\omega_{e}^2}
	- 
	i \!\!\! \sum_{\eta\neq\{\lambda,\mu\}}
	\frac{\xi_{\mu\eta\mathbf{k}}^{{p}}\xi_{\eta\lambda\mathbf{k}}^{{e}}}{ \omega_{p} + \omega_{\eta\mu\mathbf{k}} }
	\right)
	+ \{ {p}\leftrightarrow{e}\}
	\right].
\end{align}
Some further algebra shows that the second bracket is just the conjugate of the first one, so that (choosing the form of the second bracket)
\begin{align}
	\label{eq:F3}
	{\cal F}^{(3)}_{\lambda\mu\mathbf{k}} &=
	\omega_{p} 
	f_{\lambda\mu\mathbf{k}} 	
	\left|
	\left(
	\frac{\xi_{\mu\lambda\mathbf{k};{e}}^{{p}}}{\omega_{e}}
	-
	\frac{\xi_{\mu\lambda\mathbf{k}}^{{p}}
		\partial^{{e}} \omega_{\mu\lambda\mathbf{k}} }{\omega_{e}^2}
	- 
	i \!\!\! \sum_{\eta\neq\{\lambda,\mu\}}
	\frac{\xi_{\mu\eta\mathbf{k}}^{{p}}\xi_{\eta\lambda\mathbf{k}}^{{e}}}{ \omega_{p} + \omega_{\eta\mu\mathbf{k}} }
	\right)
	+ \{ {p}\leftrightarrow{e}\}
	\right|^2
\end{align}
For a two-band model, only the first two terms in the bracket are relevant.
The function ${\cal F}^{(3)}_{\lambda\mu\mathbf{k}}$ is in this form directly seen to be real-valued and thus it can be substituted into the contour integral formula \eqref{eq:alpha_isosurface} to obtain the 2PA coefficient
\begin{align}
	\beta = \frac{1}{2} \beta_{pe} \equiv \frac{1}{2} \alpha^{(3)}_{pe} 
	&= \frac{\pi}{2} \tilde{K}_3 \kappa_{D} \ 
	\sum_{\lambda\mu} 
	\oint \mathrm{d} \mathbf{k}_{\mathrm{res}}
	\frac{ {\cal F}^{(3)}_{\lambda\mu\mathbf{k}} }{
		\left\| \nabla_{\mathbf{k}} {\omega}_{\mu\lambda\mathbf{k}}\right\|}.
	\label{eq:alpha2_isosurface}
\end{align}
The dimension of $\beta$ is that of length over power. 
It is also directly obvious from these equations that $\beta_{pe}/\hbar{\omega_{p}}$ is invariant with respect to an exchange of probe and pump indices ($\{ {p}\leftrightarrow{e}\}$), which is expected, since it corresponds to the photon number absorption rate and each beam delivers one photon.
This symmetry is to be distinguished from the better known intrinsic permutation symmetry \cite{Boyd2008} of susceptibilities/conductivities, since it additionally relies on the consistency of the band-structure model. 
The reason is that we are not symmetrizing over all indices, but exclude permutations in \eqr{eq:sigma_n} contributing to  pump absorption, such as 
$\tilde{\sigma}^{(3);epep}(\omega_{e};\omega_{p},\omega_{e},-\omega_{p})$.
Interestingly, this kind of setup/investigation thus provides a check of sum rules, which are often thought to be more relevant in velocity gauge analyses.

\subsection{Implementation and local gauge fixing}
\label{ssec:implementation}

\begin{figure}[t!]
	\centering
	\includegraphics[width=7.5cm]{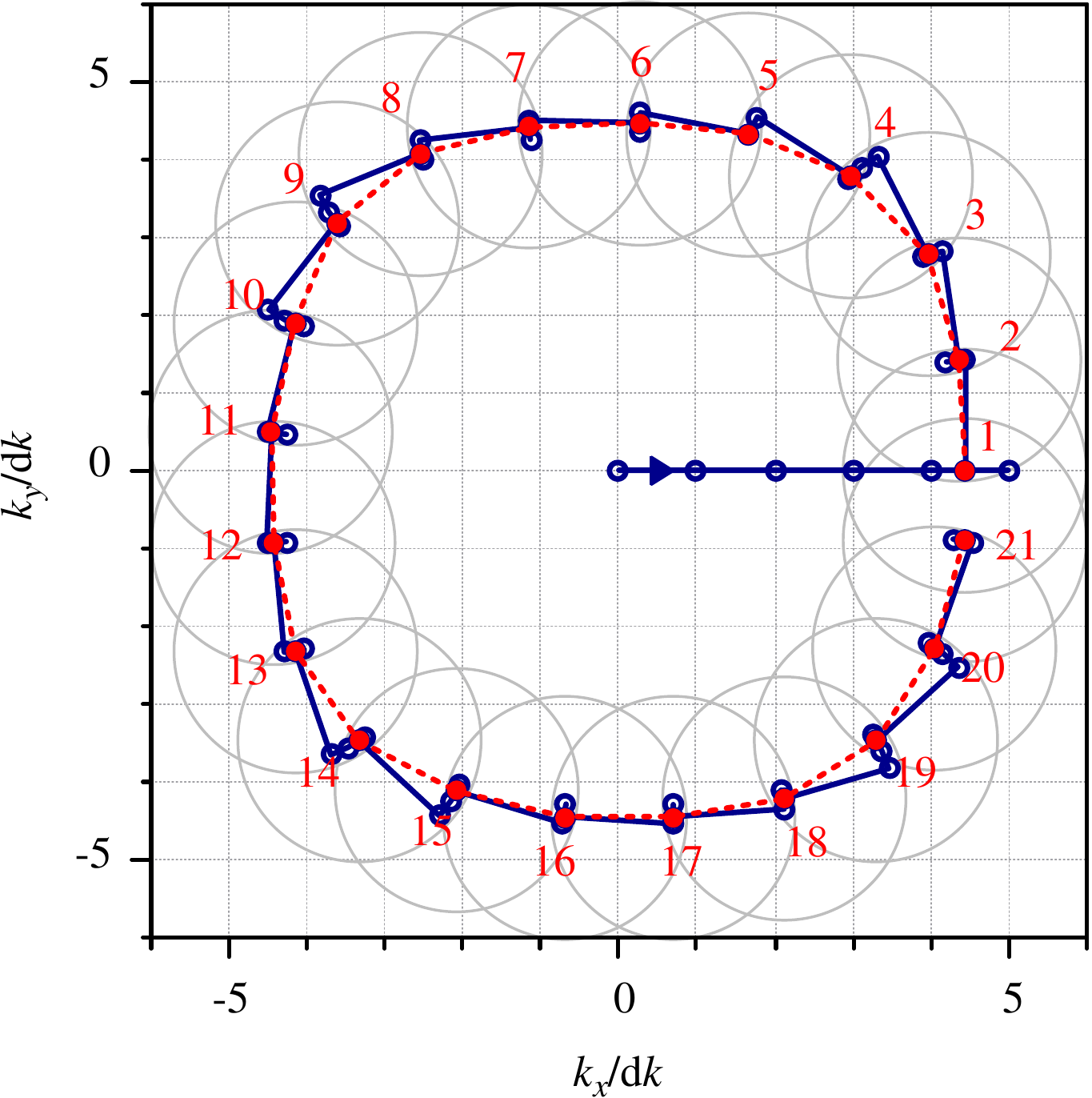}
	\caption{
		Example path (dark blue) in the algorithm for determining an ensemble of $\mathbf{k}$-points on a resonance contour line (red numbered points connected by dashed lines). }
	\label{fig:implementation}
\end{figure}
\begin{figure}[t!]
	\centering
	\includegraphics[width=7.5cm,trim=0 0 0 0]{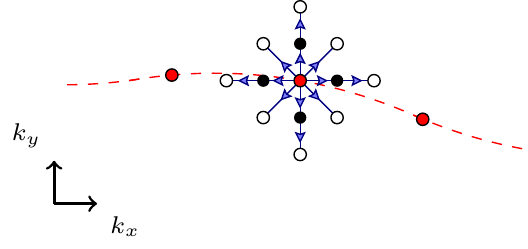}
	\caption{
		Local gauge restoration scheme to determine band-structure properties at each resonant $\mathbf{k}$-point (red circles). Eigenfunctions are required for all points of the `neighbour grid' and fixed in gauge as shown by arrows. Connection elements $\bm{\xi}_{\mathbf{k}}$ are  determined at the nearest neighbours (black circles), in order to obtain $\nabla_{\mathbf{k}}\bm{\xi}_{\mathbf{k}}$ at the central point. }
	\label{fig:localgauge}
\end{figure}

In the case of a bulk crystal, the resonance integral is further split into a $k_z$-integral (which is regularly discretized) and a resonance line integral (for each discrete value of $k_z$).
The task is now to find the 2D resonance contour and to determine the band-structure properties for each discrete point on the contour line. 

First we determine the band energies on all points of a regular $\mathbf{k}$-grid (`interpolation grid').
In order to find the resonance line for a  particular VB-CB transition, we follow a particular scheme illustrated in Fig.~\ref{fig:implementation}, where the grid (step size d$k$) is coarse for demonstration purposes. A starting point (here $k_x=k_y=0$) is required that is encircled by the searched resonance line. 
From here, we move straight along the $k_x$-axis until a resonance line is crossed. The precise position is determined by (linear/cubic) interpolation, and we also determine the normal vector $\hat{\bm{n}}_{\mathbf{k}}=\nabla_{\mathbf{k}} \omega_{\lambda\mu\mathbf{k}}/||\nabla_{\mathbf{k}} \omega_{\lambda\mu\mathbf{k}}||$ (within the $k_x,k_y$-plane) at the determined resonance point $\mathbf{k}_1$. This is the first point for the resonance line. The subsequent points $\mathbf{k}_{i+1}$ are iteratively found by moving on a circle around the point $\mathbf{k}_{i}$. The radius of this circle is about the interpolation grid step size. The start angle is chosen as the direction of the tangent vector at the previous point, i.e., the normal of the determined $\hat{\bm{n}}_{\mathbf{k}}$. The precise angle is again found using interpolation, here on a regular angle grid. 
When $|\bm{k}_i-\bm{k}_1|<k_\mathrm{circ}$ for some $i>2$, the end of the closed contour line is reached ($N\equiv i$). The first and last point ($\bm{k}_1$ and $\bm{k}_N$) are weighted by $(|\bm{k}_N-\bm{k}_1|+k_\mathrm{circ})/2$ instead of $k_\mathrm{circ}$ in order to account for the shorter distance.

The thus produced list of resonant $\mathbf{k}$-points (along with weights) is fed as input to the electronic structure program, this time determining all the required properties ($\omega_{\lambda\mathbf{k}}^{}, \bm{\xi}_{\mathbf{k}},\nabla_{\mathbf{k}}\omega_{\lambda\mathbf{k}}^{},\nabla_{\mathbf{k}}\bm{\xi}_{\mathbf{k}}$). In order to derive the connection elements and their derivatives using balanced difference quotients, we use the local neighbour grid points shown in Fig.~\ref{fig:localgauge}. Here the gauge of each point is maximally aligned to its neighbour, using the overlap matrix as in a parallel transport gauge (PTG),~\cite{Vanderbilt2018} as shown by the arrows, starting from the central (resonant) $\mathbf{k}$-point. If two bands are nearly degenerate (at the central $\mathbf{k}$-point), they are treated as a group, within eigenfunctions are mixed. Note that the derived properties are not necessarily smooth over the ensemble of resonant $\mathbf{k}$-points, so this gauge does not work for dynamical simulations, but in this analytical approach each resonant $\mathbf{k}$-point can be evaluated independently. 
Alternatively to our choice of neighbour grid, one could also use a local grid of 4$\times$4 points with the resonant $\mathbf{k}$-point in the center (not being a grid point itself), and derive the required properties by bicubic interpolation. In this scheme, a local radial gauge could also be applied.

\section{Application to specific materials}
\label{sec:applications}

\subsection{Graphene (Nearest-neighbour tight-binding model)}
\label{ssec:graphene}

In this section we apply our approach to models with analytically known eigensystem, so that no gauge fixing (like the one described in Fig.~\ref{fig:localgauge}) is needed.
For monolayer graphene, the evaluation of 1PA and 2PA is fully analytical. The numerical algorithm for finding the resonance contour lines is applied for bilayer graphene.
Some results in this section are possibly more straightforward to obtain in the velocity gauge.~\cite{Rioux2011}

\subsubsection{Graphene monolayer}
\label{sssec:Monolayer}

Here we evaluate 1PA and 2PA in intrinsic undoped graphene  over the full spectral range of the $\pi,\pi^*$-bands.
We choose the nearest-neighbour (NN) tight binding model,
$H=-\gamma_0\sum_\mathbf{R}\sum_\alpha \hat{a}_\mathbf{R}\hc \hat{b}_{\mathbf{R}+\bm{\delta}_\alpha}+\mathrm{h.c.}$,  where $\hat{a}_\mathbf{R}$ ($\hat{b}_\mathbf{R}$) is the annihilation operator for an electron on site $\mathbf{R}$ belonging to sublattice $A$ ($B$), $\bm{\delta}_\alpha$ is a NN vector, and $-\gamma_0$ is the hopping constant.
For simplicity, we neglect the overlap integral~\cite{McCann2013} between orbitals on different sites, even though this is only a good approximation in the low-energy regime.
At the same time, the model still has an improved validity range compared to the effective Hamiltonian valid near the $K$ points, which has been used previously in the evaluation of 1PA and 2PA coefficients.~\cite{Rioux2011,Yang2011,Cheng2014,Cheng2015}
Nevertheless, we are able to obtain analytical results for 1PA and 2PA coefficients. 

The parameter $\gamma_0$ is usually in the range between 2.5 and 3.0 eV; here we set it to the value $\gamma_0=3.0$ eV.
The spin index is omitted; the spin degeneracy is accounted for by a factor of $g_s = 2$ at the end.
The kernel of the Fourier transformed Hamiltonian is
\begin{align}
	{\cal H}_\mathbf{k} &= 
	-\gamma_0
	\left(\begin{array}{*{2}{c}}
		0 & s_\mathbf{k} \\ s_\mathbf{k}^* & 0
	\end{array}\right), \quad
	s_{\bm{k}} = \sum_{\alpha=1}^3 e^{-i \mathbf{k}\cdot \bm{\delta}_\alpha}.
\end{align}
The NN vectors are chosen as
\begin{align}
	\bm{\delta}_1=\delta \begin{pmatrix} 0 \\ 1 \end{pmatrix}, \ \bm{\delta}_2=\delta\begin{pmatrix} -\sqrt{3}/2 \\ -1/2 \end{pmatrix},\ \bm{\delta}_3=\delta \begin{pmatrix} \sqrt{3}/2 \\ -1/2 \end{pmatrix},
\end{align}
where $\delta=a_0/\sqrt{3}$ is the NN distance and $a_0=2.46$ \AA\ is the lattice constant.
The eigenvalues and normalized eigenvectors are
\begin{align}
	\label{eq:monolayer_ev}
	\hbar\omega_{\eta\mathbf{k}} &= 
	\sigma_\eta \gamma_0 \left| s_\mathbf{k} \right|, \quad
	\Psi_{\eta\mathbf{k}} =
	\frac{1}{\sqrt{2}}
	\left(\begin{array}{*{2}{c}}
		-\sigma_\eta s_\mathbf{k}/|s_\mathbf{k}| \\ 1
	\end{array}\right), \quad
\end{align}
with $\eta=c,v$ (which are just single band indices in this two-band model), and
 $\sigma_v=-1$ (VB) and $\sigma_c=1$ (CB). 
The model has electron-hole symmetry, and the hopping constant enters only by scaling the band energies. 
The square modulus of $s_\mathbf{k}$ is evaluated as
\begin{align}
	|s_\mathbf{k}|^2 &= 
	3+2 \left\{ \cos\left[ \mathbf{k}\cdot (\bm{\delta}_1-\bm{\delta}_2+0\bm{\delta}_3)\right] + \mathrm{c.p.} \right\}
	\\
	&=
		3+2\cos(a_0 k_x)+4\cos(a_0 k_x/2)\cos(\sqrt{3}a_0 k_y/2)
\end{align}
and vanishes at the two inequivalent $K$ points (e.g., $\mathbf{K}_{1,2}=\pm (1,0)^\mathsf{T}4\pi/3a_0$).
Here, $\mathrm{c.p.}$ stands for cyclic permutation of the NN indices.
For the connection elements we find
\begin{align}
	\bm \xi_{\eta\eta'\mathbf{k}} &=
	\sigma_\eta\sigma_{\eta'} \frac{i}{4}	
	\frac{s_\mathbf{k}^* \partial_\mathbf{k} s_\mathbf{k} - s_\mathbf{k} \partial_\mathbf{k} s_\mathbf{k}^* }{|s_\mathbf{k}|^2}
	=
	-\sigma_\eta\sigma_{\eta'}
	\frac{  \left\{  \bm{\delta}_3 \cos\left[ \mathbf{k}\cdot (\bm{\delta}_1-\bm{\delta}_2)\right] + \mathrm{c.p.} \right\} }{2|s_\mathbf{k}|^2}
	\\
	&=
	\sigma_\eta\sigma_{\eta'}
	\frac{1 }{2|s_\mathbf{k}|^2}
	\begin{pmatrix}
		\sqrt{3}\sin(a_0 k_x/2)\sin(\sqrt{3}a_0 k_y/2) \\
		-\cos(a_0 k_x)+\cos(a_0 k_x/2)\cos(\sqrt{3}a_0 k_y/2)
	\end{pmatrix}.
\end{align}
In this way, the intraband elements are just the negative of the (real) interband elements, $\xi_{vv\mathbf{k}}=\xi_{cc\mathbf{k}}=-\xi_{vc\mathbf{k}}=-\xi_{cv\mathbf{k}}$.
Around the $K$ points, the connection elements are a rotation field with a singularity at $K$.
The required $\mathbf{k}$-derivatives of $\omega_{\eta\mathbf{k}}$ and $\bm \xi_{\eta\eta'\mathbf{k}}$ are  straightforward to obtain.

\begin{figure}
	\centering
	\includegraphics[width=9cm]{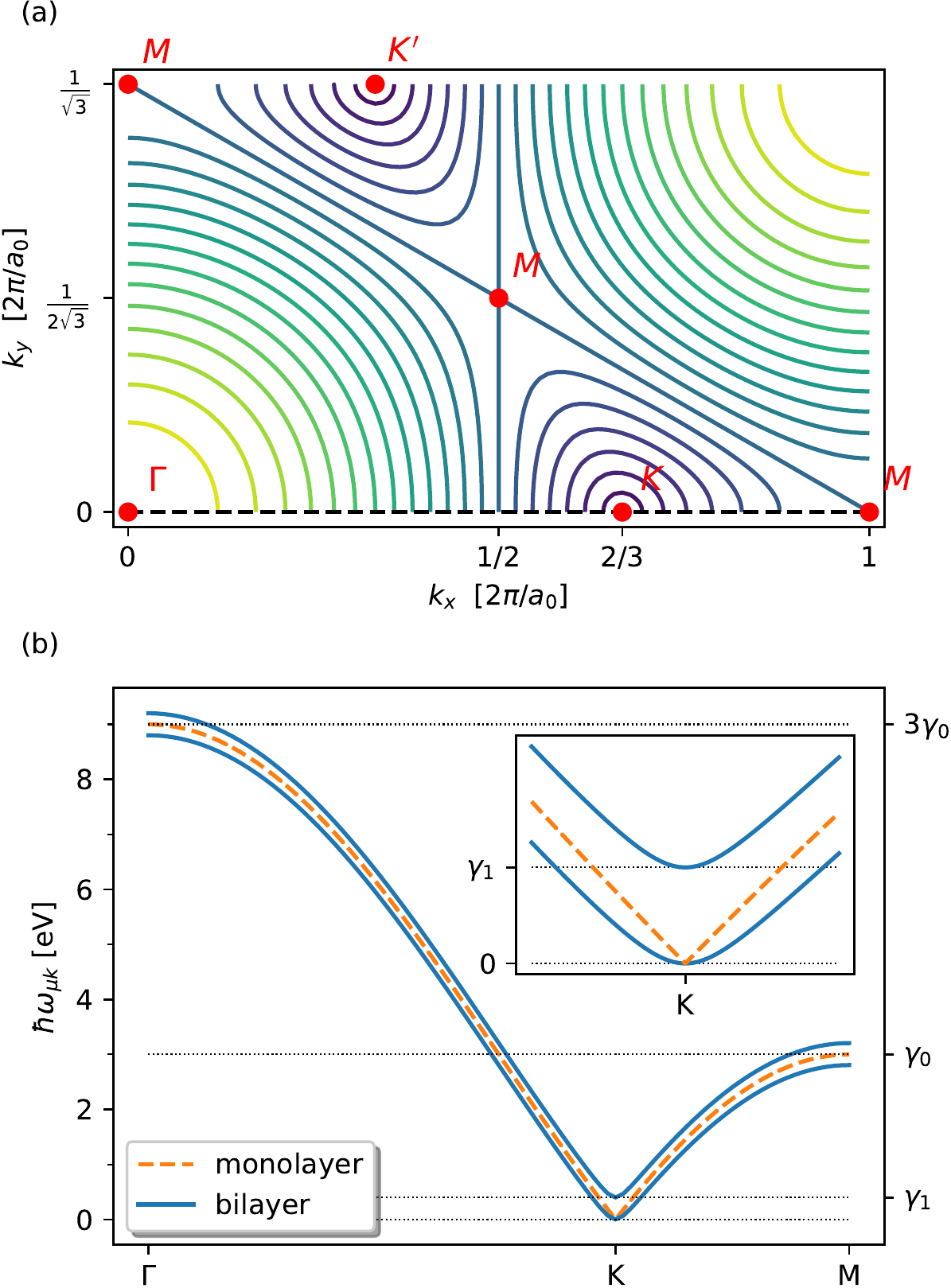}
	\caption{(a) Energy contours for the monolayer graphene VB-CB transition, $2 \gamma_0 |s_\mathbf{k}|$, cf.~\eqr{eq:monolayer_ev}. (b) The CB spectrum of monolayer (dashed) and bilayer (solid) graphene along the $k_y=0$ line, indicated by the dashed line in (a). The inset shows a zoom close to the $K$ region.
		The spectrum is electron-hole symmetric.}
	\label{fig:graphene_spectrum}
\end{figure}

Using the analytically derived structure input, it is possible to directly evaluate the $k_\perp$-integral \eqref{eq:alpha_intdkperp}, which is easier than the resonance contour integral \eqref{eq:alpha_isosurface}.
We choose $\hat{\mathbf{k}}_\parallel=\hat{\bm{y}}$, 
so we have  (setting $f_{vc\mathbf{k}}=1$ for the ground state)
\begin{align}
	\label{eq:alphan_dkx}
	\alpha^{(n)}_{p\{e\}}  
	&= \pi \tilde{K}_n\ 
	\kappa_{2} 
	\int \mathrm{d}k_x
	\sum_{k_{y,\mathrm{res}}}
	\frac{{\cal F}^{(n)}_{vc\mathbf{k}}}{\left|\partial^y {\omega}_{vc\mathbf{k}}\right|}.
\end{align}
For the following analysis it is convenient to introduce a dimensionless frequency variable
$\zeta\equiv\hbar\omega_\Sigma/2\gamma_0$.
The resonance condition $\omega_{cv\mathbf{k}}=\omega_\Sigma$ is equivalent to
\begin{align}
	\cos(\sqrt{3}a_0 k_y/2) &= 
	\frac{\zeta^2 -3-2\cos(a_0 k_x) }{4\cos(a_0 k_x/2)},
\end{align}
which, together with trigonometric identities, allows us to eliminate $k_y$ from \eqr{eq:alphan_dkx}.
For $0<\zeta<1$ the resonance contours encircle the $K$ points, while for $1<\zeta<3$ they encircle the $\Gamma$ point, see Fig.~\ref{fig:graphene_spectrum}(a).
Further using multiple-angle formulae, the dependence on $k_x$ is effectively through $u\equiv 2\cos(a_0k_x/2)$, which we substitute. 
This gives
\begin{align}
	\label{eq:alpha_graphene_I}
	\alpha^{(n)}_{p\{e\}}  &= \pi g_s \tilde{K}_n\ \kappa_{2} {\cal I}^{(n)} ,\\
	{\cal I}^{(n)}&\equiv 4
	\int_{k_{x,\mathrm{min}}}^{k_{x,\mathrm{max}}} \mathrm{d} k_x
	\frac{{\cal F}^{(n)}_{\lambda\mu\mathbf{k}}}{\left|\partial^y {\omega}_{\lambda\mu\mathbf{k}}\right|}
	= 
	4 \int_{u_\mathrm{min}}^{u_\mathrm{max}} \frac{\mathrm{d}u}{a_0\sqrt{1-u^2/4}} 
	\frac{{\cal F}^{(n)}_{\lambda\mu\mathbf{k}}}{\left|\partial^y {\omega}_{\lambda\mu\mathbf{k}}\right|},
\end{align}
where the integration ranges have been chosen as
\begin{align}
	[u_\mathrm{min},u_\mathrm{max}] = \left\{ 
	\begin{array}{*{2}{l}}
		[ -1-\zeta, -1+\zeta] , &\quad 0<\zeta<1,\\[.3em]  
		[ -1+\zeta, 2] , &\quad 1<\zeta<3. 
	\end{array} 
	\right.
\end{align}
This corresponds to the displayed sections of contour lines below the straight line connecting the $M$ points in Fig.~\ref{fig:graphene_spectrum}(a). 
The factor 4 accounts for the other sections / the other valley.

We start the evaluation with the linear absorption, $n=1$.
The linear conductivity for a hexagonal lattice is isotropic for 
normal incidence of light onto the graphene plane, since ${\sigma}^{(1)}_{xx} = {\sigma}^{(1)}_{yy}$.
The direction $\hat{\mathbf{p}}$ can thus be chosen arbitrarily in the $xy$-plane, even though the integrand in ${\cal I}^{(n)}$ will differ. Here we choose $\hat{\mathbf{p}}=\hat{\bm{x}}$, corresponding to the zigzag direction, and obtain
\begin{align}
	{\cal I}^{(1)}
	&= 
	\int_{u_\mathrm{min}}^{u_\mathrm{max}} \mathrm{d}u
	\frac{\sqrt{4-u^2}}{2\sqrt{3}\zeta^2 u^2}
	\sqrt{ 4u^2 - \left(1+u^2-\zeta^2\right)^2}.
\end{align}
The integral can be transformed to a sum of well-known elliptical integrals, which for shortness is not shown here explicitly. 
The low-frequent limit, however, is given by a constant,
\begin{align}
	\lim_{\zeta\to 0} {\cal I}^{(1)} 
	&= 
	\lim_{\zeta\to 0}
	\frac{1}{2\zeta^2}
	\int_{-1-\zeta}^{-1+\zeta} \mathrm{d}u
	\sqrt{ 4u^2 - \left(1+u^2-\zeta^2\right)^2} \\
		&= 
	\lim_{\zeta\to 0}
	\int_{-1}^{1} \mathrm{d}x
	\sqrt{ (x-\zeta^{-1})^2- \frac{1}{4}\left(\zeta^{-1} + \zeta (x-\zeta^{-1})^2 - \zeta \right)^2 }
	\\
	& = \int_{-1}^{1} \mathrm{d}x
	\sqrt{ 1-x^2 }= \frac{\pi}{2}.
	\label{eq:I1_low-freq}
\end{align}
In the second equation we substituted $u=x\zeta-1$.
The linear absorption is
\begin{align}
	\label{eq:conductivity_monolayer}
	\alpha^{(1)} &=
	K_1 \mathrm{Re}\left[{\sigma}^{(1)}_{xx}\right]	 
	=  \pi g_s K_1 \frac{e_0^2}{\hbar} \kappa_{2} {\cal I}^{(1)}.
\end{align}
If we now set $\kappa_{2} \equiv (2\pi)^{-{2}}$ and also substitute 
$K_1 = 1/(n_{0{p}}\epsilon_0 c)$, see \eqr{eq:alpha_sigma_relation}, we obtain a 2D absorption coefficient $\alpha^{(1)}_\mathrm{2D}$
and sheet conductance $\sigma^{(1)}_\mathrm{2D}$ as
\begin{align}
	\label{eq:conductivity_monolayer}
	\alpha^{(1)}_\mathrm{2D} &
	= \frac{1}{n_{0{p}}\epsilon_0 c} \sigma^{(1)}_\mathrm{2D}, \qquad
	\sigma^{(1)}_\mathrm{2D} = \frac{e_0^2}{4\hbar}\times \frac{{\cal I}^{(1)}}{\pi/2},
\end{align}
respectively.
In the low-frequent (linear dispersion) regime we substitute \eqr{eq:I1_low-freq} and obtain the universal value $\sigma^{(1)}_\mathrm{2D}=e_0^2/4\hbar$.~\cite{Nair2008} 
The 2D absorption coefficient $\alpha^{(1)}_\mathrm{2D}$ is dimensionless and, since the thin-film limit applies, corresponds to the relative intensity change $\mathrm{d}I_{p}/I_{p}$.
For a freely suspended graphene sheet, $n_{0{p}}\approx 1$ is often used, and in the low-frequent limit results in $\alpha^{(1)}_\mathrm{2D} \approx 0.0229$, in agreement with the literature.~\cite{Kuzmenko2008,Nair2008,Holovsky2015}
Obviously these low-frequent results could have been derived from the same approach but using the graphene Hamiltonian valid near $K$ points, ${\cal H}_\mathbf{k} = \hbar v_\mathrm{F} \bm{\sigma}\cdot \mathbf{k}$, with $v_\mathrm{F}=\sqrt{3}a_0 \gamma_0/2\hbar$ the Fermi velocity and $\bm{\sigma}$ the 2D vector of Pauli matrices in sublattice space.

\begin{figure}
	\centering
	\includegraphics[width=9cm]{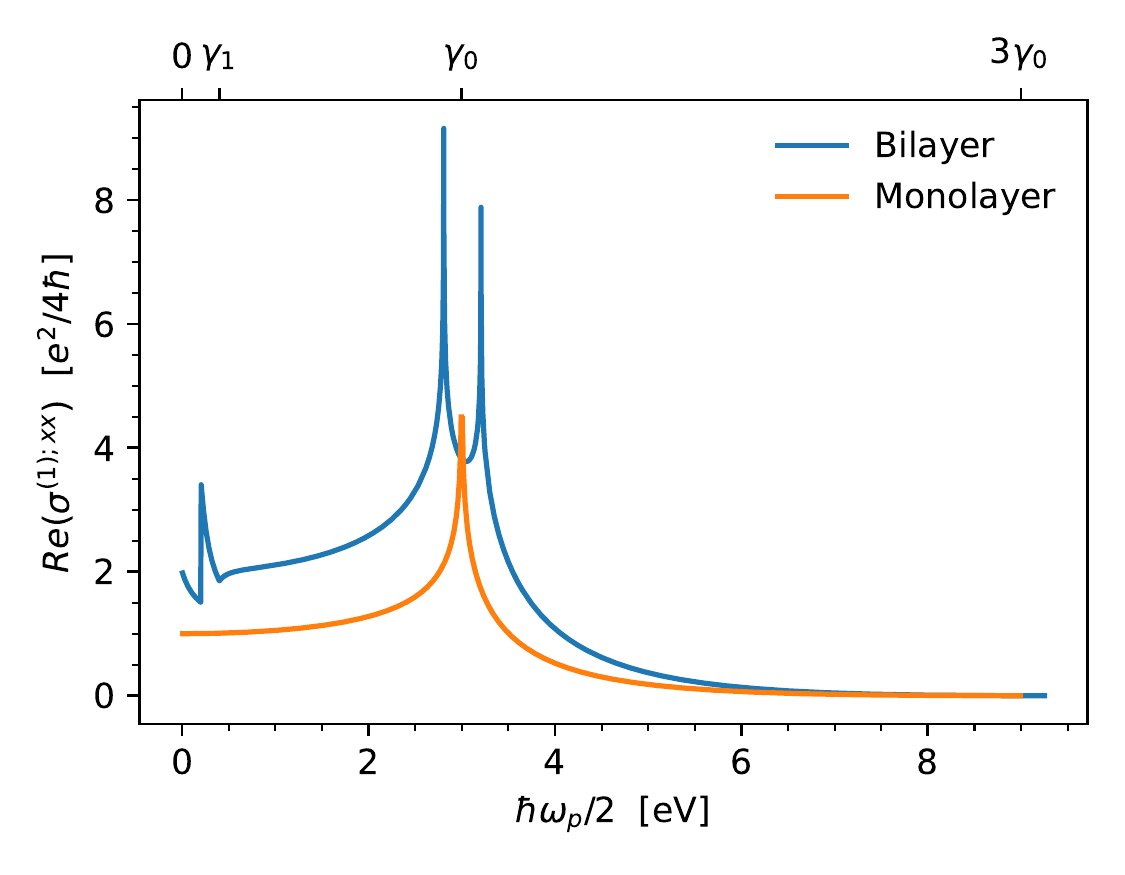}	
	\caption{Linear optical conductivity of mono- and bilayer graphene. The monolayer result corresponds to \eqr{eq:conductivity_monolayer}.}
	\label{fig:graphene_1PA}
\end{figure}

The result \eqref{eq:conductivity_monolayer}, with the ${\cal I}^{(1)}$ integral carried out numerically, is shown in Fig.~\ref{fig:graphene_1PA}.
Starting from the universal value $e_0^2/4\hbar$ in the linear dispersion regime, it increases towards a singularity at $\hbar\omega_{p}=2\gamma_0$, corresponding to the van-Hove singularity related with the saddle point at $M$. Above this peak, which becomes finite in the presence of damping, the conductivity decays and reaches zero at the maximal transition energy $\hbar\omega_{p}=6\gamma_0$.

The third-order conductivity tensor $\sigma^{(3)}$ 
of the hexagonal crystal class (point group~$6mm$) has 21 nonzero elements, of which 10 are independent.~\cite{Boyd2008}
Only the ones in the $xy$-plane are relevant here:
\begin{align}
	\sigma^{(3)}_{xxxx} &= \sigma^{(3)}_{yyyy} = \sigma^{(3)}_{xxyy} + \sigma^{(3)}_{xyxy} + \sigma^{(3)}_{xyyx} \\
	\sigma^{(3)}_{xxyy} &= \sigma^{(3)}_{yyxx}, \\
	\sigma^{(3)}_{xyxy} &= \sigma^{(3)}_{yxyx}, \\
	\sigma^{(3)}_{xyyx} &= \sigma^{(3)}_{yxxy}.
\end{align}
We consider the degenerate 2PA only, $\omega_{p}=\omega_{e}\equiv\omega$, for which intrinsic permutation symmetry further implies $\sigma^{(3)}_{xyxy} = \sigma^{(3)}_{xxyy}$.
The pump-probe absorption coefficient for an arbitrary linearly polarized pump and probe ($\hat{\mathbf{p}}=\hat{\bm{x}}\cos(\theta_{p})+\hat{\bm{y}}\sin(\theta_{p})$) is
\begin{align}
	\label{eq:alpha2_sigma3}
	\beta = \frac{1}{2} \beta_{pe} 
	&= \frac{1}{2\epsilon_0^2 n_0^2 c^2} \left( \sigma'^{(3)}_{xxxx} \cos^2(\theta_{p}-\theta_{e}) + \sigma'^{(3)}_{xyxy}  \sin^2(\theta_{p}-\theta_{e}) \right),
\end{align}
where single (double) prime denotes real (imaginary) part.
This pump-probe response is isotropic, i.e., unchanged by a rotation of the graphene sheet with respect to the normal axis. 
An evaluation of the co- and cross-polarized 2PA is obviously sufficient to determine the complete third-order response, which we do here. 
With one valence- and one conduction band in this model, the third term in the bracket in \eqr{eq:F3} vanishes. Since both intraband elements are identical ($\bm \xi_{vv\mathbf{k}}=\bm \xi_{cc\mathbf{k}}$), the generalized derivative becomes the usual one.
In the co-polarized case, denoted by the subscript $\parallel$, we obtain 
\begin{align}
	{\cal I}^{(3)}_\parallel
	&= 
	\int_{u_\mathrm{min}}^{u_\mathrm{max}} \mathrm{d}u
	\frac{a_0^2 \hbar^2}{4\sqrt{3}\gamma_0^2 u^4 \zeta^8 \sqrt{4-u^2}}
	\sqrt{ 4u^2 - \left(1+u^2-\zeta^2\right)^2} 
	\nonumber\\
	&\qquad\qquad
	\times \left(8+2 u^4-8 \zeta ^2+u^2 \left(\zeta ^2-10\right)\right)^2.
\end{align}
Here we have chosen $\hat{\mathbf{p}}=\hat{\mathbf{e}}=\hat{\bm{x}}$; as above, for other directions, the integrand differs but the integral remains the same.
As for 1PA, the integral can be reduced to a sum of elliptical integrals, but we choose to evaluate it numerically.
In the low-frequent regime it scales as $\zeta^{-4}$ with a constant prefactor given by
\begin{align}
	\lim_{\zeta\to 0} \zeta^4{\cal I}^{(3)}_\parallel
	&= 
	\lim_{\zeta\to 0}
	\frac{a_0^2\hbar^2}{12 \gamma_0^2 \zeta^4}
	\int_{-1-\zeta}^{-1+\zeta} \mathrm{d}u
	\sqrt{ 4u^2 - \left(1+u^2-\zeta^2\right)^2} 
	\nonumber\\
	&\qquad
	\times\left(8+2 u^4-8 \zeta ^2+u^2 \left(\zeta ^2-10\right)\right)^2 \\
	&= 
	\lim_{\zeta\to 0}
	\frac{a_0^2\hbar^2}{12 \gamma_0^2 }
	\int_{-1}^{1} \mathrm{d}x
	\sqrt{ 4(x-\zeta^{-1})^2-\left(\zeta^{-1} + \zeta ( (x-\zeta^{-1})^2 - 1 ) \right)^2 } \nonumber\\
	&\qquad
	\times \left(\frac{8}{\zeta}+2\zeta^3 (x-\zeta^{-1})^4-8 \zeta+\zeta(x-\zeta^{-1})^2 \left(\zeta^2-10\right)\right)^2 \\
	& = 
	\frac{24 a_0^2\hbar^2}{\gamma_0^2 }
	\int_{-1}^{1} \mathrm{d}x
	\sqrt{ 1-x^2 } x^2
	= 	
	\frac{3\pi a_0^2\hbar^2}{\gamma_0^2 },
	\label{eq:I2_lowfreq}
\end{align}
with the same substitution $u=x\zeta-1$ as for 1PA.
The explicit expression for the cross-polarized result ${\cal I}^{(3)}_\perp$ is skipped here. We note, however, that its low-frequent limit is identical with the co-polarized case.

From \eqr{eq:alpha_graphene_I}, the degenerate 2PA coefficient is 
\begin{align}
	\beta_\mathrm{2D} = \frac{1}{2} \alpha^{(3)}_{pe} &= g_s \frac{\pi}{2} \tilde{K}_3\ \kappa_{2} {\cal I}^{(3)}
	=
	\frac{e_0^4}{8\pi n_{0{p}}n_{0{e}} \hbar^3}
	\left(\frac{1}{\epsilon_0 c}\right)^2 {\cal I}^{(3)}.
\end{align}
In the low-frequent regime, substituting \eqr{eq:I2_lowfreq} we find the scaling
$\beta_\mathrm{2D} = \alpha^{(3)}_{pe} /2 = \bar{K}/\omega^4$ with
the roughly constant prefactor 
\begin{align}
	\bar{K} &\equiv
	\frac{e_0^4}{ n_{0{p}}n_{0{e}} \hbar^5}
	\left(\frac{1}{\epsilon_0 c}\right)^2 \frac{3 a_0^2 \gamma_0^2}{8}
	=
	\frac{1}{ 2 n_{0{p}}n_{0{e}} \hbar^3}
	\left(\frac{v_\mathrm{F}e_0^2}{ \epsilon_0 c }\right)^2 .
	\label{eq:2PA_graphene_monolayer_Dirac}
\end{align}
Such scaling of the 2PA coefficient with the excitation frequency has been found earlier \cite{Rioux2011,Yang2011}.

\begin{figure}
	\centering
		\includegraphics[width=9cm]{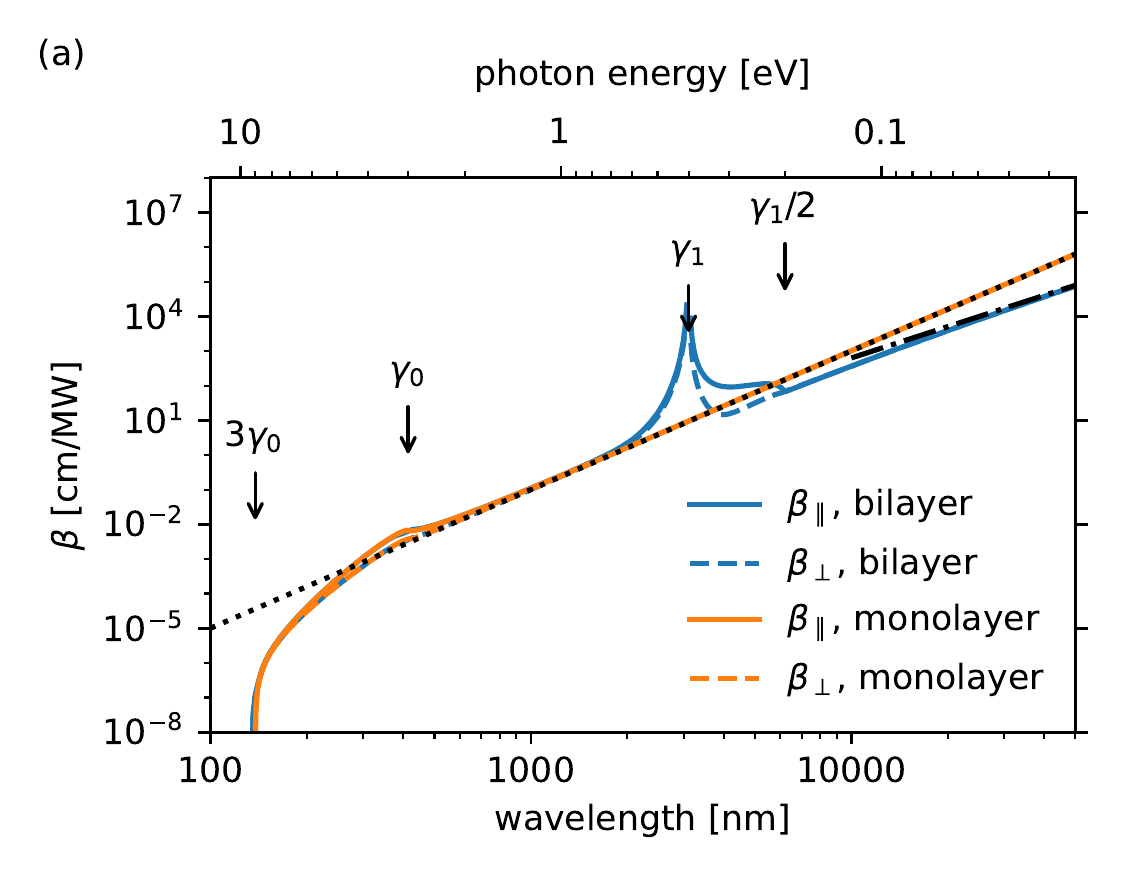}	
		\vskip-.5em
		\includegraphics[width=9cm]{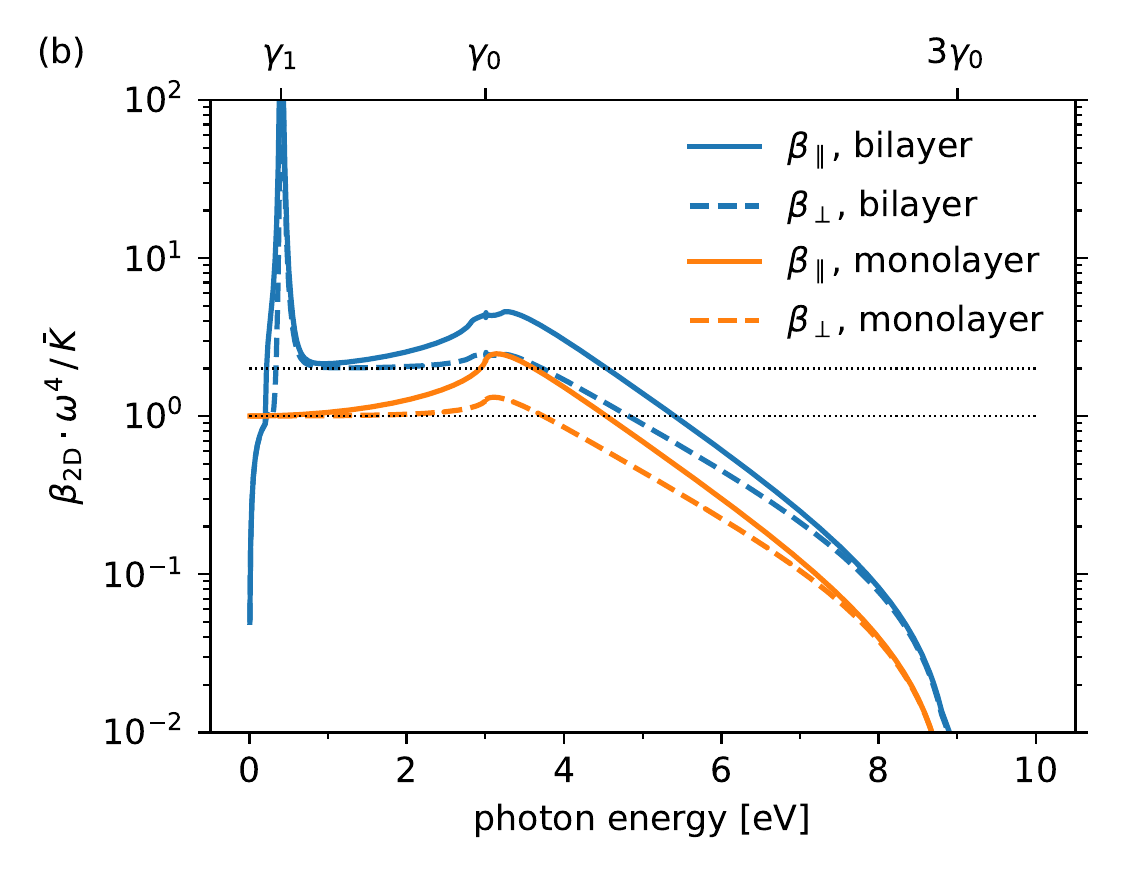}	
	\caption{Degenerate 2PA coefficients of mono- and bilayer graphene for co- and cross-polarized excitation. 
		These are isotropic, i.e., independent of the light polarization with respect to the crystallographic axes. 
	In (a), $\beta$ is shown as a function of photon wavelength, while in (b) the same data is shown using a different scaling and as a function of energy.
	In (a), the dotted line is $\beta = (\bar{K} / L_\mathrm{c}) \omega^{-4}$,
	the dash-dotted line is $\sim \omega^{-3}$. 
	In (b), the 2D absorption coefficient (i.e., not divided by $L_\mathrm{c}$), normalized by the long-wavelength result for the monolayer, is shown. The horizontal lines correspond to the values 1 and 2.}
	\label{fig:graphene_2PA}
\end{figure}

Multiplying the absorption coefficient $\beta_\mathrm{2D}$, with units [area/power], by the pump intensity~$I_{e}$ yields half the relative absorption $\mathrm{d}I_{p}/I_{p}$ of the probe by the 2D material due to the presence of the pump. 
Dividing $\beta_\mathrm{2D}$ by a thickness, i.e., setting $\kappa_{2}=(2\pi)^{-2}L_\mathrm{c}^{-1}$, yields a  bulk 2PA coefficient.
For the thickness we substitute the layer spacing in graphite, $L_\mathrm{c}=3.3$\AA.
The result for $\beta$ is shown in Fig.~\ref{fig:graphene_2PA}~(a), along with $\beta_\mathrm{2D} \omega^4 / \bar{K}$ in Fig.~\ref{fig:graphene_2PA}~(b).
Here we used a frequency-independent value $n_0=\sqrt{3}$.~\cite{Yang2011}
The characteristic scaling with $\omega^{-4}$ holds for photon energies smaller than $\gamma_0$, where co- and cross-polarized 2PA is identical. In the vicinity of $\hbar\omega=\gamma_0$ there is some enhancement. In this regime the co-polarized 2PA becomes about twice as large as the cross-polarized one.
At higher energies the 2PA coefficients decrease and drop to zero at the band edge, $\hbar\omega=3\gamma_0$.

\subsubsection{Graphene bilayer}
\label{sssec:Bilayer}

A similar evaluation is now carried out 
for Bernal-stacked (AB-stacked) bilayer graphene. The analogous model has four bands and serves as an example of a multiband model. 
We consider only NN interlayer hopping characterized by the constant  $\gamma_1$. The bilayer Hamiltonian is
\begin{align}
	{\cal H}_\mathbf{k} &= 
	\left(\begin{array}{*{4}{c}}
		0 & -\gamma_0 s_\mathbf{k} & 0 & \gamma_1 \\ -\gamma_0 s_\mathbf{k}^* & 0 & 0 & 0 \\ 0 & 0 & 0 & -\gamma_0 s_\mathbf{k} \\ \gamma_1 & 0 & -\gamma_0 s_\mathbf{k}^* & 0
	\end{array}\right).
\end{align}
For simplicity, we again set the overlap matrix to unity and keep the same value for $\gamma_0$ as for monolayer in Sec.~\ref{sssec:Monolayer}, while in reality it is slightly larger.
For $\gamma_1$ we use a value in the typical range, namely 0.4 eV.~\cite{Kuzmenko2009,McCann2013} 
As shown in Fig.~\ref{fig:graphene_spectrum}(b), the two VBs/CBs in this model follow the monolayer bands with an offset of $\gamma_1/2$ in nearly the entire Brillouin zone. In the vicinity of $K$ all four bands are parabolic instead of linear.

An analytical solution (up to the final $k_\parallel$-integration) as for monolayer graphene might be possible, but leads to lengthy expressions. Instead we evaluate the absorption coefficients numerically, using the contour integral method. The eigenvalues and -vectors of ${\cal H}_\mathbf{k}$ are entered as analytical expressions so that no gauge fixing is needed. The derivatives ($\xi_{\mathbf{k}}, \partial \xi_{\mathbf{k}}, \partial \omega_{\mathbf{k}}$) are computed as balanced difference quotients (see Sec.~\ref{ssec:implementation}). 

Even though the crystal structure of bilayer graphene belongs to a different point group, its relevant tensor properties are identical with the monolayer ones. Therefore the response in first and third order is isotropic, so that we may again choose $\hat{\mathbf{p}}=\hat{\bm{x}}$ and $\hat{\mathbf{e}}=\hat{\bm{x}}$ ($\hat{\mathbf{e}}=\hat{\bm{y}}$) for the co- (cross-) polarized case.
The bilayer 1PA and 2PA results are shown along with the monolayer results in Figs.~\ref{fig:graphene_1PA} and~\ref{fig:graphene_2PA}. To obtain a bulk absorption coefficient with units [length/power], we use the same dielectric constant as for monolayer, $n_0=\sqrt{3}$, and the doubled thickness, $L_\mathrm{c}=6.6$\AA.

For long wavelengths 
the linear conductivity, depicted in Fig.~\ref{fig:graphene_1PA}, is on average twice as strong as that of a monolayer. However, a singularity along with a jump exists at $\hbar\omega_{p}=\gamma_1$. 
There is no singularity at $\hbar\omega_{p}=2\gamma_0$ as for the monolayer, but instead two singularities occur at $2\gamma_0\pm \gamma_1$.
Apart from that, the 1PA spectra of mono- and bilayer look similar as expected from the band structure.

By contrast, the bilayer degenerate 2PA coefficient, plotted in Fig.~\ref{fig:graphene_2PA}, differs from the monolayer result already in the low-frequent regime, where it scales with $\omega^{-3}$ for $\hbar\omega \ll \gamma_1$.~\cite{Rioux2011,Yang2011}
In the range $\gamma_1/2 < \hbar\omega < \gamma_1$ the co- and cross-polarized 2PA strongly differ.
At $\hbar\omega=\gamma_1$ there is a resonance peak, which in our case of absent scattering is a singularity.
These findings are in agreement with the results of Refs.~\cite{Rioux2011,Yang2011}.
For higher photon energies, the bilayer 2PA (bulk) coefficients are very similar to the monolayer results. 
This means that $\beta_\mathrm{2D}$ of the bilayer is about twice as large as of the monolayer, which can be seen in Fig.~\ref{fig:graphene_2PA}(b). 
We thus cannot confirm the conclusion of Ref.~\cite{Yang2011} that the bilayer 2PA is two orders of magnitudes stronger than that of the monolayer in the VIS/NIR regime, however, this finding seems to be also in disagreement with the results of Ref.~\cite{Rioux2011}.

\subsection{Zincblende semiconductors (${\bm {k\cdot}}{\bm p}$ theory)}
\label{ssec:Zincblende}

In this section we apply our analytical approach to more complicated band-structure models. 
In particular, we consider Zincblende semiconductors described by well-established ${\bm {k\cdot}}{\bm p}$ models.
First we analyze the polarization dependence of $\beta$, which can be compared with the tensor symmetry properties for the given crystal class, and later we compare $\beta$ with results from dynamical simulations. 

\subsubsection{Polarization dependence of degenerate 2PA}
\label{sssec:pol_dep}

\begin{figure}[b!]
	\centering
	\includegraphics[height=18em,trim=4em 0 0 0]
	{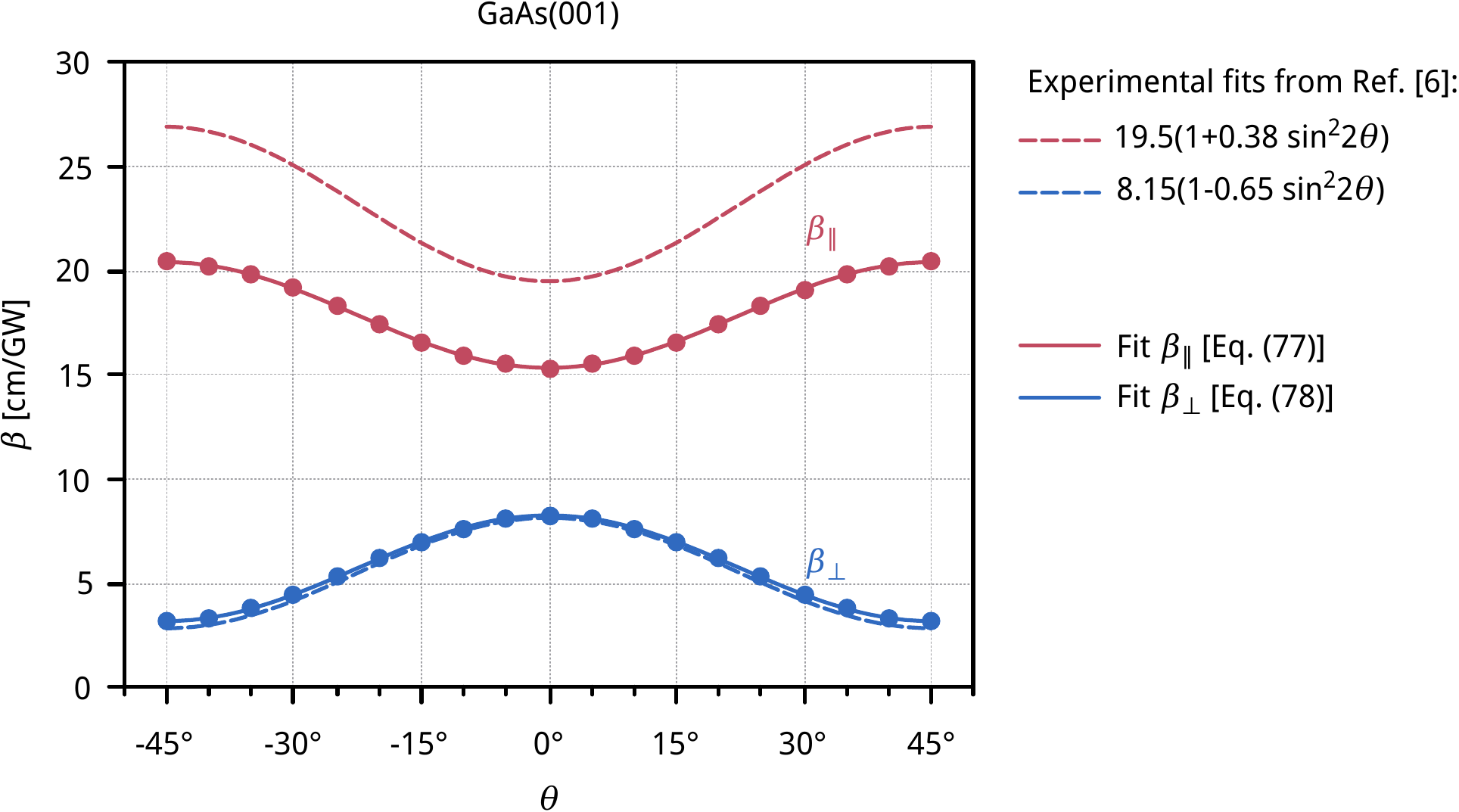}
	\begin{textblock}{10}(4.1,-1.2)
		\footnotesize
		Fit parameters:\\
		\includegraphics[height=5em,trim=0 0 0 0]
		{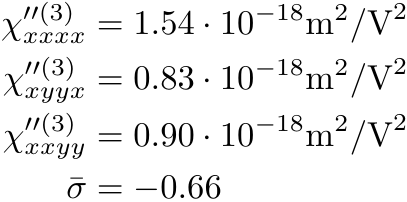}
	\end{textblock}	
	\caption{
		2PA coefficient of Zincblende semiconductors as a function of the probe polarization direction ($\theta_{p}$). The pump is co-polarized ($\theta_{e}\!=\!\theta_{p}$) or cross-polarized ($\theta_{e}\!=\!\theta_{p}\!+\!\pi/2$); the angle $\theta$ may be interpreted here as either $\theta_{p}$, or, to match with Ref.~\cite{Dvorak1994}, $\theta_{e}$. 
	}
	\label{fig:polarplot_alpha2}
\end{figure}
A set of 2PA measurements has been carried out and analyzed by Dvorak \textit{et al}~\cite{Dvorak1994} with bulk Zincblende semiconductors.
From this set we consider the polarization-dependent degenerate 2PA coefficient of GaAs(001) (at room temperature), with $\hbar\omega_{p}=\hbar\omega_{e}=1.3$eV. 
The band-structure input is obtained from the 30-band ${\bm {k\cdot}}{\bm p}$-model,~\cite{Richard2004} with the band gap set to $\hbar\omega_\mathrm{g} = 1.42$eV.

The third-order susceptibility tensor $\chi^{(3)}$ 
of Zincblende structures (crystal group $\bar{4}3m$) has 21 nonzero elements, of which 4 are independent.~\cite{Boyd2008}
Since we choose the propagation direction of both pulses in $z$ (in crystal coordinates, i.e., [001]), the polarization unit vectors lie in the $xy$-plane and we need to consider only elements therein, namely
\begin{align}
	\sigma^{(3)}_{xxxx} &= \sigma^{(3)}_{yyyy}, \\
	\sigma^{(3)}_{xxyy} &= \sigma^{(3)}_{yyxx}, \\
	\sigma^{(3)}_{xyxy} &= \sigma^{(3)}_{yxyx}, \\
	\sigma^{(3)}_{xyyx} &= \sigma^{(3)}_{yxxy}.
\end{align}
In the degenerate case studied here, we further have $\sigma^{(3)}_{xyxy} = \sigma^{(3)}_{xxyy}$ due to intrinsic permutation symmetry.
The angle between the pump polarization and the $x$-axis ([100]) is denoted by $\theta$. 
The probe is chosen as either co- or cross-polarized.
Following Ref.~\cite{Dvorak1994}, the corresponding 2PA coefficients are found as
\begin{align}
	\label{eq:alpha2_chi3_co}
	\beta_\parallel &= \frac{1}{2} \beta_{pe}^\parallel = \frac{1}{2\epsilon_0^2 n_0^2 c^2} \sigma'^{(3)}_{xxxx}
	\left[ 1 + \bar{\sigma} \left( \cos^4\theta + \sin^4\theta - 1 \right) \right], \\
	\label{eq:alpha2_chi3_cross}
	\beta_\perp &= \frac{1}{2} \beta_{pe}^\perp = \frac{1}{2\epsilon_0^2 n_0^2 c^2} \left( 2 \bar{\sigma} \sigma'^{(3)}_{xxxx} \cos^2\theta \sin^2\theta + \sigma'^{(3)}_{xyxy} \right),
\end{align}
where  the anisotropy parameter is given by
\begin{equation}\label{key}
	\bar{\sigma} = 1-\left( \sigma'^{(3)}_{xyyx} + 2\sigma'^{(3)}_{xyxy}\right) / \sigma'^{(3)}_{xxxx}.
\end{equation}
Remember that the Cartesian indices are ordered as ${\sigma}^{(n)}_{\beta\alpha_1\dots\alpha_n}$, 
not ${\sigma}^{(n)}_{\beta\alpha_n\dots\alpha_1}$ as in Ref.~\cite{Dvorak1994}.

Figure~\ref{fig:polarplot_alpha2} shows the calculated 2PA coefficients as scatter plot (circles), along with 
fits using Eqs.~\eqref{eq:alpha2_chi3_co} and~\eqref{eq:alpha2_chi3_cross} by solid lines.
The agreement of the fit curves with the data is excellent, 
meaning that the model is in perfect agreement with the crystal symmetry. 
The parameters extracted from the fit to our numerical data are listed next to the plot in Fig.~\ref{fig:polarplot_alpha2}. 
The first and last parameter have been obtained from $\beta_\parallel$, the second and third from $\beta_\perp$.
The dashed lines are the fits to experimental data from Ref.~\cite{Dvorak1994}; the experimental data itself is not shown. 
The magnitudes of the individual $\chi^{(3)}$-tensor elements agree with the experiment~\cite{Dvorak1994}, at least they are within the uncertainty ranges.
The anisotropy parameter, even though in rough agreement, is slightly too high compared with the experimental value of $-0.76 \pm 0.08$.
In the plot in Fig.~\ref{fig:polarplot_alpha2}, our cross-polarized result can be seen to be nearly identical with Ref.~\cite{Dvorak1994}, while the co-polarized curve is somewhat smaller than the experimental one but still in rather good agreement.

\subsubsection{Dependence of 2PA on the non-degeneracy parameter}

Here we carry out a comparison with dynamical simulations of non-degenerate 2PA in bulk GaAs and ZnSe.
For ZnSe these simulations have been published in a recent joint experimental and theoretical study~\cite{KraussKodytek2021}, and we start with this material.
The dynamical simulations are just numerical solutions of the perburbatively expanded SBE~\eqref{eq:SBE_length-gauge} for finite pulses. 
There, dephasing and relaxation processes have been neglected as it is done here. Gaussian pulses with FWHM of $\sim30$ fs are shorter than the experimental ones but sufficiently long so that the absorption spectrum does not change significantly for longer pulses.
The incidence of laser pulses is (approximately) normal to the sample surface, so the label `(100)-sample' refers to light propagation direction $\hat{\bm{k}}_{p}( \approx \hat{\bm{k}}_{e}) \parallel [100]$. The individual polarization settings are as follows (illustrated in Ref.~\cite{KraussKodytek2021}):
\begin{itemize}
	\item (100)-sample: $\hat{\mathbf{p}} \parallel [011]$, $\hat{\mathbf{e}}_\perp \parallel [0\bar{1}1]$
	\item (110)-sample: $\hat{\mathbf{p}} \parallel [1\bar{1}0]$, $\hat{\mathbf{e}}_\perp \parallel [001]$.
\end{itemize}
Here $\hat{\mathbf{e}}_\perp$ is defined as the direction perpendicular to both $\hat{\mathbf{p}}$ and $\hat{\bm{k}}_{p}$. 
For co-polarized 2PA, $\hat{\mathbf{e}}=\hat{\mathbf{p}}$, and for cross-polarized 2PA, $\hat{\mathbf{e}}=\hat{\mathbf{e}}_\perp$.
Both theoretical approaches are based on the same eight-band Kane model,~\cite{Winkler2003} where for convenience the (100)-sample is replaced by the equivalent setting  $\hat{\bm{k}}_{p} \parallel [001]$, $\hat{\mathbf{p}} \parallel [110]$, $\hat{\mathbf{e}}_\perp \parallel [\bar{1}10]$.
Also, for the (110)-sample we perform a rotation of the coordinate system within the ${\bm {k\cdot}}{\bm p}$-model, such that the $z$-axis corresponds to [110]. 
As in Ref.~\cite{KraussKodytek2021} we model the 
frequency-dependent refractive index $n_0(\omega)$ for ZnSe by
$n_0^2(\omega) = A + {B\lambda^2}/({\lambda^2-C^2})$,
where $\lambda$ is the wavelength in micrometers, and the parameters 
have the values $A=4.00$, $B=1.90$, and $C^2=0.113$.~\cite{Marple1964}

\begin{figure}
	\centering
	\includegraphics[width=9cm]{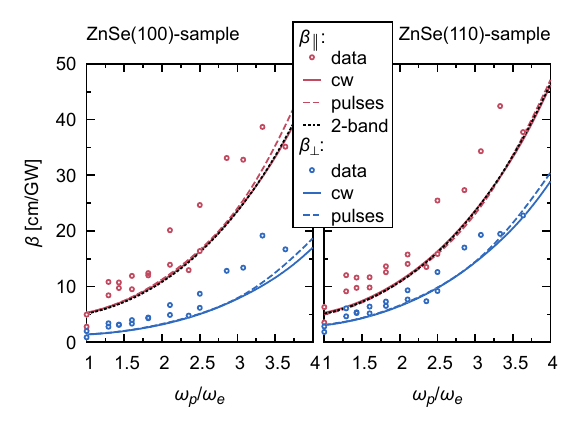}
	\caption{Dependence of bulk ZnSe 2PA coefficients $\beta_\perp$,$\beta_\parallel$ on the non-degeneracy parameter. The sum frequency corresponds to $\hbar\omega_\Sigma=3.098$eV. The experimental data labeled `data' is the one shown in Ref.~\cite{KraussKodytek2021}; no error bars are available. The polarization settings for each sample are also defined therein. The curves labeled `cw' are obtained from the analytical approach presented in this work, using the same 8-band ${\bm {k\cdot}}{\bm p}$ model as in Ref.~\cite{KraussKodytek2021}. 
	The curves labeled `pulses' are the results from dynamical simulations already shown in Ref.~\cite{KraussKodytek2021}, but corrected by a small post-processing mistake, a constant factor $\pi/3$. All curves and data are for room temperature. The curve labeled `2-band' is a fit of $\beta_\parallel \sim n_{0p}^{-1} n_{0e}^{-1} \omega_{p}^{-3} \omega_{e}^{-4}$ to the `cw'-curve.}
	\label{fig:ZnSe_nondeg2PA}
\end{figure}

\begin{figure}
	\centering
	\includegraphics[width=9cm]{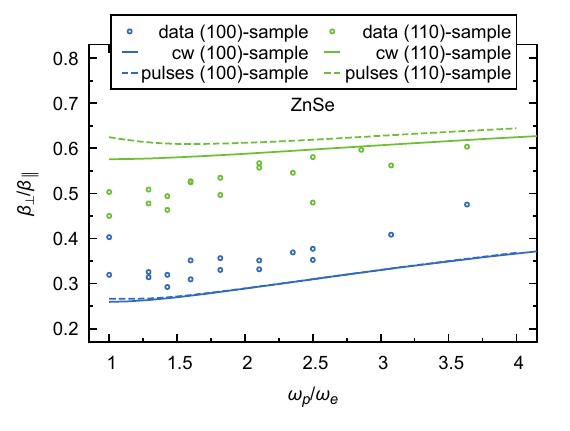}
	\caption{Same data as in Fig.~\ref{fig:ZnSe_nondeg2PA} but plotting the ratio of cross- and co-polarized 2PA coefficients $\beta_\perp/\beta_\parallel$. See the caption of Fig.~\ref{fig:ZnSe_nondeg2PA} for details.}
	\label{fig:ZnSe_nondeg2PA_ratio}
\end{figure}

Figure~\ref{fig:ZnSe_nondeg2PA} shows the 2PA coefficient $\beta = \beta_{pe}/2$ in the pump-probe setting with fixed frequency sum $\omega_\Sigma=3.098$eV$/\hbar$.
The results of the dynamical simulations can be seen to be nearly identical to the analytical approach. The analytical approach describes a steady state response to monochromatic plane waves, since we evaluate the 2PA result~\eqref{eq:alpha2_isosurface} at the discrete carrier frequencies only. 
In Fig.~\ref{fig:ZnSe_nondeg2PA_ratio} the corresponding ratio $\beta_\perp/\beta_\parallel$ is plotted from the same data. Here one can distinguish a slightly better fit of the analytical approach to the measurement in the case of the (110)-sample.

\begin{figure}
	\centering
	\includegraphics[width=9cm]{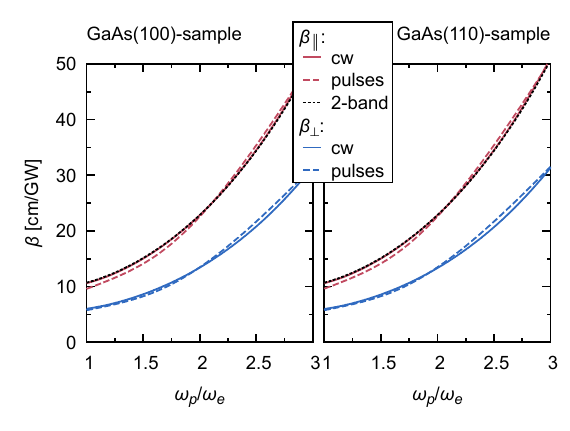}
	\caption{Comparison of the analytical approach with dynamical simulations as in Fig.~\ref{fig:ZnSe_nondeg2PA}, for bulk GaAs with $\hbar\omega_\Sigma=1.5694$eV. The `pulses' curves have some uncertainty and are spline fits to only three discrete data points per curve, since the computational expense is large.}
	\label{fig:GaAs_nondeg2PA}
\end{figure}

In Fig.~\ref{fig:GaAs_nondeg2PA} a similar comparison is carried out for bulk GaAs, also modeled by the eight-band Kane model.~\cite{Winkler2003}
The four polarization settings are (illustrated in Ref.~\cite{KraussKodytek2021a}):
\begin{itemize}
	\item (100)-sample: $\hat{\mathbf{p}} \parallel [010]$, $\hat{\mathbf{e}}_\perp \parallel [001]$
	\item (110)-sample: $\hat{\mathbf{p}} \parallel [001]$, $\hat{\mathbf{e}}_\perp \parallel [1\bar{1}0]$.
\end{itemize}
Again, the (100)-sample is replaced by an equivalent setting  $\hat{\bm{k}}_{p} \parallel [001]$, $\hat{\mathbf{p}} \parallel [100]$, $\hat{\mathbf{e}}_\perp \parallel [010]$.
Since the variation of refractive index $n_0(\omega)$ in the considered light frequency range ($\sim 0.4 - 1.2$eV) is small, we have  set it to a constant value of 3.4 for simplicity.
The sum frequency corresponds to $\hbar\omega_\Sigma=1.5694$eV. Even though the Gaussian pulses have a duration of only $\sim9$ fs (FWHM), there is very good agreement between both approaches. 
Corresponding experimental data can be found in Ref.~\cite{KraussKodytek2021a} and qualitatively agrees with our theory, both in order of magnitude and the relative dependence on $\omega_{p}/\omega_{e}$. 
Surprisingly, the two sample orientations / polarization settings give nearly identical theoretical curves for each $\beta_\parallel$ and $\beta_\perp$.
This is not quite the case in the experiment~\cite{KraussKodytek2021a}, where for the (100)-sample the overall ratio $\beta_\perp/\beta_\parallel$ remains at about $1/3$. On the other hand, for the (110)-sample, both in experiment and theory this ratio starts at around $1/2$ and slightly increases for larger $\omega_{p}/\omega_{e}$.

\section{Conclusions}
\label{sec:conclusions}

We have investigated multi-photon absorption within a pump-probe setting by semiconductors and graphene in the independent-particle approximation. 
The basis of our numerical evaluations is a rather compact analytical expression for the 
2PA coefficient, which is expressed as a contour integral over the resonance isosurfaces in $\mathbf{k}$-space. In principle, similar expressions could be derived for $\ell>2$. 
Comparing Eqs.~\eqref{eq:F1} and \eqref{eq:F3}, we note that the number of terms within the square modulus of ${\cal F}^{(n)}_{\lambda\mu\mathbf{k}}$ just corresponds to $n$, so it might be that 3PA can be expressed by just five terms (some of which are sums over bands).

In order to verify the correctness of this approach,  
in the evaluation we have focused on materials that have been analyzed previously by other means, both experimentally and theoretically.
The analytical result for the 1PA and 2PA coefficients of monolayer (intrinsic) graphene in the nearest-neighbour tight-binding approximation could be compared with the literature in the low-frequent regime.
The analysis for bulk Zincblende semiconductor has been based on ${\bm {k\cdot}}{\bm p}$ theory.
Here we have shown that the results are in good agreement with dynamical simulations of the 2PA of finite-length pulses. 
For sufficiently small spectral width of the pulses, the dynamical simulation is directly comparable with the steady-state response at the carrier frequencies as we have done here, instead of integrating over their spectra. 
It should be possible to implement our algorithm in DFT programs, which would open up its application to a wider range of materials.

\section*{Acknowledgements}

We acknowledge financial support from the Guangdong Province Science and Technology Major Project (Future functional materials under extreme conditions - 2021B0301030005).

%

\appendix

\section{Relation between optical conductivity and absorption coefficient}
\label{sec:absorption_pref}

In this appendix we derive the relation between the optical conductivity of particular order $(n)$ and the corresponding absorption coefficient, taking care of obtaining the correct prefactors. 
For simplicity we do this for plane wave (pump and probe) beams, which have a sharp carrier frequency, so we start by setting
\begin{align}
	\label{eq:E-cw}
	\bm{E}_\omega &= \bm{E}_{\omega,p} + \bm{E}_{\omega,e} \\
	\bm{E}_{\omega,p} &= 2\pi \delta(\omega-\omega_{p}) \pmb{\cal E}_{p} + 
	                     2\pi \delta(\omega+\omega_{p}) \pmb{\cal E}_{p}^* 
\end{align}
and analogously for $\bm{E}_{\omega,e}$.
The Fourier transform, \eqr{eq:E-FT}, is given by
\begin{align}
	\bm{E}(\bm{r},t) &= \bm{E}_{p}(\bm{r},t) + \bm{E}_{e}(\bm{r},t)  \\
	\label{eq:E_p_time}
	\bm{E}_{p}(\bm{r},t) &= \pmb{\cal E}_{p} e^{i(\bm{k}_{p}\bm{r}-\omega_{p} t)} + \mathrm{c.c.} 
\end{align}
and analogously for $\bm{E}_{e}(\bm{r},t)$. 
Here, as for the perturbative quantities $\rho_{\mathbf{k}}^{(n)}$ and $\bm{J}^{(n)}$ below, we explicitly included the spatial dependence, which has been omitted in Sec.~\ref{sec:formalism}.
Note that the real field amplitudes are $2|\pmb{\cal E}_{p}|$ and  $2|\pmb{\cal E}_{e}|$ for the probe and pump, respectively. The intensities are thus
\begin{align}
	\label{eq:intensity}
	I_{p} = 2 \epsilon_0 n_{0p} c |\pmb{\cal E}_{p}|^2
\end{align}
and analogously for $I_{e}$.
The following subsections correspond to two alternative derivations leading to the same result.

\subsection{$\langle\bm{J}\cdot \bm{E}\rangle$-method}
$\ell$PA is determined by the $\ell$-th power of the intensity,
\begin{align}
	(I_{p} + I_{e})^\ell = I_{p}^\ell + \ell I_{p}^{\ell-1} I_{e} + \dots + \ell I_{p} I_{e}^{\ell-1} + I_{e}^\ell.
\end{align}
For the weak-probe/strong-pump absorption considered here, only the second last term is relevant. 
The rate of (pump-induced) energy absorption from the probe beam is 
\begin{align}
	\langle \bm{J}^{(n)}\cdot \bm{E}_{p} \rangle &= 
	\bm{J}^{(n)(+1|0)}\cdot \pmb{\cal E}_{p}^* e^{-i(\bm{k}_{p}\bm{r}-\omega_{p} t)} + \nonumber\\
	&\ \quad \bm{J}^{(n)(-1|0)}\cdot \pmb{\cal E}_{p} e^{i(\bm{k}_{p}\bm{r}-\omega_{p} t)} \\
	&= 2\, \mathrm{Re} \left( \bm{J}^{(n)(+1|0)}\cdot \pmb{\cal E}_{p}^* e^{-i(\bm{k}_{p}\bm{r}-\omega_{p} t)} \right)
	\\ &
	= 2\, \mathrm{Re} \left( J_{p}^{(n)(+1|0)}\cdot {\cal E}_{p}^* e^{-i(\bm{k}_{p}\bm{r}-\omega_{p} t)} \right),
\end{align}
with ${\cal E}_p = \pmb{\cal E}_{p} \cdot \hat{\mathbf{p}}$ etc.
Here, after carrying out the spatial average $\langle \dots \rangle$ over a macroscopic regime, where we employ the slowly varying amplitude approximation, only the current components co- or counter-propagating with the probe beam remain; these have the oscillatory dependence $exp[{\pm i(\bm{k}_{p}\bm{r}-\omega_{p} t)}]$ denoted by the superscript $(\pm 1|0)$.
We obtain the absorption coefficient by normalizing by the intensities,
\begin{align}
	\label{eq:alpha_n_JE}
	\alpha^{(n)}_{p\{e\}} = \ell \alpha^{(n)} 
	= \frac{2\, \mathrm{Re} \left( J_{p}^{(n)(+1|0)}\cdot {\cal E}_{p}^* e^{-i(\bm{k}_{p}\bm{r}-\omega_{p} t)} \right) }{I_{p} I_{e}^{\ell-1}}.
\end{align}
The factor $\ell$ is just the prefactor from the relevant intensity product term above. 
At the same time, it is the number of relevant permutations of field products leading to $\bm{J}^{(n)(+1|0)}$; 
for example, for 2PA ($n=3$), the combinations are ${\cal E}_p {\cal E}_e {\cal E}_e^*$ and ${\cal E}_e {\cal E}_p {\cal E}_e^*$.~\cite{Dvorak1994}

Now (see \eqr{eq:Jn}),
\begin{align}
	\label{eq:Jn_cw}
	J_{p}^{(n)(+1|0)}(\bm{r},t) &= -\frac{e_0}{L^3} \sum_{\mathbf{k}} \mathrm{Tr} \left( v_{\mathbf{k}}^{p} \rho_{\mathbf{k}}^{(n)(+1|0)}(\bm{r},t) \right),
\end{align}
and for the plane wave beams (see \eqr{eq:sol-form})
\begin{align}
	\label{eq:rho_cw}
	\rho_{\mathbf{k}}^{(n)(+1|0)}(\bm{r},t) &
	=
	\left(\frac{e_0}{\hbar} \right)^n
	 e^{i(\bm{k}_{p}\bm{r}-\omega_{p} t)}
	{\cal E}_p |{\cal E}_e|^{n-1} \bar{\cal P}_{\mathbf{k}}^{(n)}
\end{align}
with $\bar{\cal P}_{\mathbf{k}}^{(n)}$ as defined in Sec.~\ref{ssec:Absorption}.
Thus, \eqr{eq:alpha_n_JE} becomes
\begin{align}
	\alpha^{(n)}_{p\{e\}} = \ell \alpha^{(n)} 
	= - 2 \frac{e_0}{L^3} \left(\frac{e_0}{\hbar} \right)^n
	\frac{ \sum_{\mathbf{k}} \mathrm{Re}\left[ \mathrm{Tr} \left( v_{\mathbf{k}}^{p} \bar{\cal P}_{\mathbf{k}}^{(n)} \right) \right]}{ (2 \epsilon_0 c)^\ell n_{0p} n_{0e}^{\ell-1}}
\end{align}
which corresponds to \eqr{eq:alpha-ell}. Finally using the symmetrized form of the conductivity~\eqref{eq:sigma-n} we return to \eqr{eq:alpha_sigma_relation}:
\begin{align}
	\alpha^{(n)}_{p\{e\}} &
	=
	\frac{ 2}{(2 \epsilon_0 c)^\ell } \frac{1}{n_{0p} n_{0e}^{\ell-1}}  \mathrm{Re} \left[\sigma_\mathrm{eff}^{(n)}\right].
\end{align}

\subsection{Nonlinear wave equation method}

This alternative derivation is based on Ref.~\cite{Dvorak1994} and the Supplementary Material of Ref.~\cite{Furey2021}.
The nonlinear wave equation is
\begin{align}
	\left( \nabla^2 - \frac{n_{0p}^2}{c^2} \frac{\partial^2}{\partial t^2} \right) \bm{E}_{p} (\bm{r},t)
	=
	\frac{1}{\epsilon_0 c^2} \frac{\partial^2}{\partial t^2} \bm{P}^{(n)}(\bm{r},t).
\end{align}
Using the plane-wave basis, and setting $\hat{\bm{k}}_{p} = \hat{\bm{z}}$,
\begin{align}
	\left( \frac{\partial^2}{\partial z^2} - \frac{n_{0p}^2}{c^2} \frac{\partial^2}{\partial t^2} \right) {E}_{p} 
	=
	\frac{1}{\epsilon_0 c^2} \frac{\partial^2}{\partial t^2} {P}_{p}^{(n)}
	=
	\frac{1}{\epsilon_0 c^2} \frac{\partial}{\partial t} {J}_{p}^{(n)}.
\end{align}
Here ${J}_{p}^{(n)} = {J}_{p}^{(n)(+1|0)} + {J}_{p}^{(n)(-1|0)}$ is the nonlinear current density driving the optical response.
The forward- and backward propagating terms in the wave equation can be compared separately. 
Since they are the complex conjugate of each other, we continue explicitly with the forward-propagating part only, for which we find
\begin{align}
	\left( \frac{\partial^2}{\partial z^2} - \frac{n_{0p}^2}{c^2} \frac{\partial^2}{\partial t^2} \right) {\cal E}_{p} e^{i(k_{p}z-\omega_{p} t)}
	=
	\frac{1}{\epsilon_0 c^2} \frac{\partial}{\partial t} {J}_{p}^{(n)(+1|0)}.
\end{align}
The left-hand side can thus be written
\begin{align}
	\left( \frac{\partial^2 {\cal E}_{p} }{\partial z^2} + 2 i k_{p} \frac{\partial {\cal E}_{p} }{\partial z} - k_{p}^2 {\cal E}_{p} + \frac{n_{0p}^2 \omega_{p}^2}{c^2} {\cal E}_{p}  \right) e^{i(k_{p}z-\omega_{p} t)}. 
\end{align}
In the slowly-varying envelope approximation the first term in the bracket is neglected and the third and fourth terms cancel.
This gives
\begin{align}
	2 i k_{p} \frac{\partial {\cal E}_{p} }{\partial z} e^{i(k_{p}z-\omega_{p} t)} 
	=
	\frac{1}{\epsilon_0 c^2} \frac{\partial}{\partial t} {J}_{p}^{(n)(+1|0)},
\end{align}
where we can cancel the oscillatory terms to obtain  the reduced wave equation (with ${J}_{p}^{(n)(+1|0)} = {\cal J}_{p} e^{i(k_{p}z-\omega_{p} t)} $)
\begin{align}
	2 i k_{p} \frac{\partial {\cal E}_{p} }{\partial z} 
	=
	- \frac{ i \omega_{p}}{\epsilon_0 c^2} {\cal J}_{p} 
\end{align}
or
\begin{align}
	\frac{\partial {\cal E}_{p} }{\partial z} 
	=
	- \frac{ 1}{2 n_{0p} \epsilon_0 c} {\cal J}_{p} .
\end{align}

Analogously to \eqr{eq:rho_cw}, the current (\eqr{eq:J_sigma_E}), projected onto $\hat{\mathbf{p}}$ and including relevant permutations, is
\begin{align}
	\label{eq:J_cw}
	J_{p}^{(n)(+1|0)}(z, t) &=
	e^{i(k_{p}z-\omega_{p} t)}
	{\cal E}_p |{\cal E}_e|^{n-1} \sigma_\mathrm{eff}^{(n)}. 
\end{align}
We thus have
\begin{align}
	\frac{\partial {\cal E}_{p} }{\partial z} 
	=
	- \frac{ 1}{2 n_{0p} \epsilon_0 c} {\cal E}_p |{\cal E}_e|^{n-1} \sigma_\mathrm{eff}^{(n)}
\end{align}
and correspondingly for the complex conjugate.
This allows us to write the derivative of the intensity~\eqref{eq:intensity} as
\begin{align}
	\frac{\partial I_{p} }{\partial z}  &
	=
	2 n_{0p} \epsilon_0 c \left( {\cal E}_p  \frac{\partial {\cal E}_{p}^* }{\partial z} + \mathrm{c.c.} \right) \\ &
	=
	- 2 |{\cal E}_p|^2 |{\cal E}_e|^{n-1} \mathrm{Re} \left[\sigma_\mathrm{eff}^{(n)}\right].
\end{align}
On the other hand, the absorption coefficient is defined as
\begin{align}
	\frac{\partial I_{p} }{\partial z}  &
	=
	\alpha^{(n)}_{p\{e\}} I_{p} I_{e}^{\ell-1},
\end{align}
so we get that
\begin{align}
	\alpha^{(n)}_{p\{e\}} &
	=
	\frac{ 2}{(2 \epsilon_0 c)^\ell } \frac{1}{n_{0p} n_{0e}^{\ell-1}}  \mathrm{Re} \left[\sigma_\mathrm{eff}^{(n)}\right],
\end{align}
which corresponds to \eqr{eq:alpha_sigma_relation}.

If instead of single plane waves, we have a spectrum,
\begin{align}
	\label{eq:E-FT_k}
	\bm{E}(z,t)= \int \frac{\mathrm{d}\omega}{2\pi} { \bm{E}}_\omega e^{i(k(\omega) z - \omega t)}
\end{align}
with $k(\omega)=n_0(\omega)\omega/c$, the derivation goes analogously, since each frequency component can be treated independently, similar to the splitting into forward/backward modes above.

\section{$\mathbf{k}_\perp$-integral}
\label{sec:transform-k-integral}

Here we show that\footnote{The band indices $\mu\lambda$ will be skipped in this appendix.}
\begin{equation}\label{eq:surface-contour}
	\int \mathrm{d}^{{D}-1}\mathbf{k}_\perp
	\sum_{k_{\parallel,\mathrm{res}}}
	\left| \frac{\partial {\omega}_{\mu\lambda\mathbf{k}}}{\partial k_\parallel}	\right|^{-1} \dots
	=
	\oint \mathrm{d} \mathbf{k}_{\mathrm{res}}
	\left\| \nabla_{\mathbf{k}} {\omega}_{\mu\lambda\mathbf{k}}\right\|^{-1}
	\dots
\end{equation}

We first give a proof for ${D}=2$. Denoting the parallel and perpendicular axes by $y$ and $x$, respectively, we can write for one specific resonance path (there are two of them if the isoline is convex):
\begin{align}
	\label{eq:transform-k-integral}
	\int \mathrm{d} k_x = \int \mathrm{d} s \frac{\mathrm{d} k_x}{\mathrm{d} s} = \int \mathrm{d} s\ \hat{t}_{\mathbf{k},x} = 
	\int \mathrm{d} s\ \frac{\left| \frac{\partial {\omega}_{\mathbf{k}}}{\partial k_y}	\right|}{\left\| \nabla_{\mathbf{k}}^{xy} {\omega}_{\mathbf{k}}\right\|},
\end{align}
where $s$ is the arc length along the path and $\hat{\bm{t}}_{\mathbf{k}}$ is the tangential unit vector on the path. In the last equation, $\hat{\bm{t}}_{\mathbf{k}}$ has been expressed by rotating the normal unit vector $\hat{\bm{n}}_{\mathbf{k}}$ by $\pi/2$,
\begin{align}
	\label{eq:tangent-vec}
	\hat{\bm{t}}_{\mathbf{k}} = \pm 
	\left(\begin{array}{*{2}{c}}
		0 & 1 \\ -1 & 0
	\end{array}\right)
	\hat{\bm{n}}_{\mathbf{k}}, \quad
	\hat{\bm{n}}_{\mathbf{k}} = 
	\frac{\nabla_{\mathbf{k}}^{xy} {\omega}_{\mathbf{k}}}{\left\| \nabla_{\mathbf{k}}^{xy} {\omega}_{\mathbf{k}}\right\|}, \quad
	\nabla_{\mathbf{k}}^{xy} \equiv 
	\left(\begin{array}{*{1}{c}}
		\partial/\partial{k_x} \\ \partial/\partial{k_y}
	\end{array}\right)
\end{align}
where $\pm = \mathrm{sgn}\left(\partial {\omega}_{\mathbf{k}}/\partial k_y\right)$ in order to keep the direction of the integration ($\hat{t}_{\mathbf{k},x}>0$). In the last expression of (\ref{eq:transform-k-integral}) the direction doesn't matter any more, and we can also put together the two resonance paths to form a closed contour. 
Then we directly arrive at \eqr{eq:surface-contour},
where for ${D}=2$ the gradient is understood as $\nabla_{\mathbf{k}}=\nabla_{\mathbf{k}}^{xy}$.

Now we show that \eqr{eq:surface-contour} also holds for ${D}=3$, where it reads (choosing $\hat{\mathbf{k}}_\parallel=\hat{\bm{z}}$)
\begin{align}
	\int \mathrm{d}k_x \mathrm{d}k_y
	\sum_{k_{z,\mathrm{res}}}
	\left| \frac{\partial {\omega}_{\mu\lambda\mathbf{k}}}{\partial k_\parallel}	\right|^{-1} \dots
	&=
	\oint \mathrm{d} \mathbf{k}_{\mathrm{res}}
	\left\| \nabla_{\mathbf{k}} {\omega}_{\mu\lambda\mathbf{k}}\right\|^{-1}
	\dots
	\label{eq:trafo-k-integral_3D}
\end{align}
We start from the right-hand side, where we first divide the closed surface into `layers' (in the $z$-direction) of individual surfaces $S_{\mathrm{res}}$, so that each $S_{\mathrm{res}}$ is describable by a function $k_{z,\mathrm{res}}(k_x,k_y)$ in a (simply connected) subspace of the $k_x,k_y$-plane.
Then we express the surface integral as a plane 2D integral using the appropriate surface element,
\begin{align}
	\oint \mathrm{d} \mathbf{k}_{\mathrm{res}}
	&= 
	\sum_{k_{z,\mathrm{res}}}
	\int\!\!\!\!\int \mathrm{d} S_{\mathrm{res}} \\
	&= 
	\sum_{k_{z,\mathrm{res}}}
	\int\!\!\!\!\int \mathrm{d} k_x \mathrm{d} k_y
	\left\|
	\frac{\partial\mathbf{k}_\mathrm{res}}{\partial k_x} \times
	\frac{\partial\mathbf{k}_\mathrm{res}}{\partial k_y}\right\|, 
\end{align}
with $\mathbf{k}_\mathrm{res}^\mathsf{T} = (k_x, k_y, k_{z,\mathrm{res}}(k_x,k_y) )$.
Now we use that
\begin{align}
	\frac{\partial k_{z,\mathrm{res}}}{\partial k_x} &=
	\frac{\hat{t}^{xz}_z}{\hat{t}^{xz}_x}
	= \frac{\partial_{k_x} {\omega}_{\mathbf{k}}}{\partial_{k_z} {\omega}_{\mathbf{k}}}, 
	\quad
	\frac{\partial k_{z,\mathrm{res}}}{\partial k_y} =
	\frac{\hat{t}^{yz}_z}{\hat{t}^{yz}_y}
	= \frac{\partial_{k_y} {\omega}_{\mathbf{k}}}{\partial_{k_z} {\omega}_{\mathbf{k}}},
\end{align}
where, for instance, $\hat{\bm{t}}^{xz}$ is the tangent vector to the resonance contour in the $xz$-plane, which is again expressed according to \eqr{eq:tangent-vec} (the sign $\pm$ now doesn't matter from the beginning). Therefore
\begin{align}
	&\left\|
	\frac{\partial\mathbf{k}_\mathrm{res}}{\partial k_x} \times
	\frac{\partial\mathbf{k}_\mathrm{res}}{\partial k_y}
	\right\|
	=
	\left\|
	\left(\begin{array}{*{1}{c}}
		1 \\ 0 \\ \hat{t}^{xz}_z / \hat{t}^{xz}_x
	\end{array}\right)
	\times
	\left(\begin{array}{*{1}{c}}
		0 \\ 1 \\ \hat{t}^{yz}_z / \hat{t}^{yz}_y
	\end{array}\right)
	\right\| \\
	&=
	\left\|
	\left(\begin{array}{*{1}{c}}
		- \hat{t}^{xz}_z / \hat{t}^{xz}_x \\
		- \hat{t}^{yz}_z / \hat{t}^{yz}_y \\
		1
	\end{array}\right)
	\right\|
	=
	\frac{\left\| \nabla_{\mathbf{k}} {\omega}_{\mathbf{k}}\right\|}{|\partial_{k_z} {\omega}_{\mathbf{k}}|}
\end{align}
Applying these transformations to the right-hand side of~\eqr{eq:trafo-k-integral_3D} completes the proof for ${D}=3$.

\end{document}